\documentclass[a4paper,12pt]{article}

\usepackage{amsmath,amssymb,amscd,mathtools,array,empheq}
\usepackage[utf8]{inputenc}
\usepackage{lmodern,newtxtext,newtxmath}
\usepackage[a4paper,tmargin=3truecm,bmargin=3truecm,hmargin=1.6truecm]{geometry}
\usepackage{graphicx}
\usepackage{color}
\usepackage[dvipsnames]{xcolor}
\usepackage[english]{babel}
\usepackage{bm}
\usepackage{cite}
\usepackage[colorlinks=true, pdfstartview=FitV, linkcolor=blue, citecolor=blue, urlcolor=blue]{hyperref}
\usepackage{hyperref}
\hypersetup{colorlinks=true, urlcolor=blue, citecolor=blue, linktoc=page}

\DeclareMathOperator{\Tr}{Tr}

\DeclareMathOperator{\Ad}{Ad}

\newcommand{\R}{\mathbb{R}}

\newcommand{\SL}{\mathrm{SL}}
\newcommand{\PSL}{\mathrm{PSL}}
\newcommand{\PGL}{\mathrm{PGL}}
\newcommand{\CO}{\mathrm{CO}}
\newcommand{\GL}{\mathrm{GL}}
\newcommand{\SO}{\mathrm{SO}}

\newcommand{\so}{\mathfrak{so}}
\newcommand{\gl}{\mathfrak{gl}}

\newcommand{\Ric}{\mathrm{Ric}}

\newcommand{\ie}{\textit{i.e.} }
\newcommand{\eg}{\textit{e.g.} }
\newcommand{\half}{\frac{1}{2}}

\newcommand{\cC}{\mathcal{C}}

\newcommand{\cP}{\mathcal{P}}

\newcommand{\lieg}{\mathfrak{g}}
\newcommand{\womega}{\widetilde{\omega}}
\newcommand{\wOmega}{\widetilde{\Omega}}
\newcommand{\we}{\widetilde{e}}
\newcommand{\walpha}{\widetilde{\alpha}}

\usepackage{scalerel}
\let\savewidetilde\widetilde
\def\widetilde#1{%
 \ThisStyle{\savewidetilde{\phantom{\SavedStyle#1}}%
  \setbox0=\hbox{$\SavedStyle#1$}\kern-\wd0#1}}

\title{Conformal and projective tractors from 2-frame bundles by the dressing field method}

\author{S. Lazzarini\footnote{mailto: lazzarini@cpt.univ-mrs.fr} \, and \, L. Marsot\footnote{mailto: loic.marsot@univ-amu.fr}\\[1.em]
{\normalsize Centre de Physique Théorique}\\
{\normalsize Aix Marseille Univ, Université de Toulon, CNRS, CPT, Marseille, France.}%
} 

\date{{\footnotesize (\today)}}

\begin{document}

\maketitle

\smallskip

\begin{abstract}
This paper is devoted to the study of conformal and projective structures, and especially their connections, in the language of 2-frames, or $G$-structures of 2nd-order. While their normal Cartan connections are well-known, we use the dressing field method to obtain their local, mostly gauge invariant, connections. Lastly, we recall how to obtain the tractor bundle from conformal structures, and then show that we can define a projective tractor bundle using the same kind of construction.
\end{abstract}


\section{Introduction}

Conformal and projective structures date back to Cartan \cite{Cartan23b,Cartan24,Cartan37} and have since been studied by a few authors. In the 1960s, Kobayashi-Nagano \cite{KobayashiN64} worked on projective structures, while Ogiue \cite{Ogiue67} worked on conformal Cartan connections. Both geometries were later summarized in a standard textbook by Kobayashi \cite{Kobayashi72}. All these works use the language of 2-frame bundles, or $G$-structures. They have attracted some interest for physical models in the past, see e.g. \cite{Harnad:1976yu,Harnad:1976hg,Harnad:1978it}.

A more modern approach was given by Sharpe \cite{Sharpe96}, using the matrix representation instead of 2-frames. Yet, somehow, his approach is dissatisfying as we ``lose'' the fine control that was given by the 2-frame representation. For connections, this materializes as the fact that their components are identified \textit{a posteriori} instead of \textit{a priori} in the previous approach.

Often, for example when doing Physics, one considers not the (global) Cartan connection on a bundle, but a connection that is local and stripped of its gauge symmetries. For instance, going from the first connection to the latter kind is a well-known procedure on the usual principal bundle $P(M, H)$ where $M$ is locally $G/H$, with $G$ the Poincaré group $\SO(n-1,1) \ltimes \R^n$ and $H$ the Lorentz group $\SO(n-1,1)$. Indeed, one can take the spin part of the Cartan connection and recover the Christoffel symbols through a gauge-like transformation, see for example in  \cite{GockelerS87}. 

While this process is rather straightforward on $P(M, \SO(n-1,1))$, we wish to study different geometries, such as conformal and projective structures. In the case of conformal structures, such process has been somewhat developed in the literature. For instance in Sharpe's work \cite{Sharpe96}, it takes the form of a gauge fixing to get rid of one of the components of the Cartan connection. In the works of Kobayashi-Nagano \cite{KobayashiN64} and Ogiue \cite{Ogiue67} it takes the form of the choice of a special (or ``natural'')
section to express the connection through. While working, these processes are not satisfactory, as they seem rather arbitrary. We will use in this paper a third approach yielding equivalent results, but which is constructive.

Indeed, quite recently a method has been developed \cite{TheseCarina,TheseJordan} to constructively reduce the gauge symmetries of a connection, named the \emph{dressing field method}. With the help of dressing fields, one is able to render the Cartan connection, as well as the curvature, invariant under certain gauge transformations. For instance \cite{FournelFLM14}, take the Cartan connection $\omega$ on $P(M, \SO(n-1,1))$, with $\omega = \left(\begin{matrix}
\bar{\omega} & \theta \\
0 & 0
\end{matrix}\right)$ where $\bar{\omega}$ is the spin connection and $\theta$ is the \emph{soldering} form. After the dressing field process, one is left with the local Levi-Civita connection $\varpi_0 = \left(\begin{matrix}
\Gamma^\lambda_{\mu\nu} dx^\nu & dx^\lambda \\
0 & 0
\end{matrix}\right)$, where $\Gamma$ are the Christoffel symbols, which is inert under the gauge group $\mathcal{SO}$.

\smallskip
In this paper, our intent is to give a modern and precise look at the construction of (local) conformal and projective connections, by first computing the normal Cartan connections, and then look at them locally, through the dressing field process. The language of 2-frame bundles, which we will describe in section~\ref{section_geometry}, is better suited to understand such connections as we will see. Though the connections will be parametrized by the language of 2-frame bundles, we will keep the matrix representation, as it makes computations for the dressing field method much easier, and as it is the representation most often used in practice. This method is applied in section~\ref{section_dressing}, after a quick review in~\ref{ss:review_dressing} of the basics. In the end, we will recover the usual conformal connection as described, for example, in Sharpe's work \cite{Sharpe96}, in section \ref{ss:conf}. We will also find the local, and mostly gauge invariant as we will see, expression of the Levi-Civita connection for projective structures, in section \ref{ss:proj}.

Lastly, there is a direct link between the connections expressed in local coordinates and tractors bundles \cite{BaileyEG94,GoverM17}, which can be made through the dressing field method. See \cite{CurryG14} for an overview of Tractors, and \cite{AttardF18} for the construction of tractor bundles with the dressing field method in the case of conformal structures. We will show in the end that the dressing field method can also be used to define projective tractor bundles, not only conformal ones.

\section{The geometry of conformal and projective structures}
\label{section_geometry}

\subsection{2-frame bundles}

\label{section_2-frames}

Let us recall the definition of 2-frame bundles, for instance as given in \cite{Ogiue67}. Recall that 2-frame bundles, and in general $k$-frame bundles, aim to generalize the principal bundle of linear frames \cite{Kobayashi61}, \ie $P(M, \GL(n, \R))$, using jet theory. The 2-frame bundle $P^2(M) = P(M, G^2(n))$ is a principal bundle made up of a base manifold $M$, with $n = \dim M$\footnote{We consider the cases $n \ge 3$ in this paper.}, and of structural group $G^2(n)$. Let us now construct this group. First, consider a 2nd order jet $j^2_0(f)$ at $0$ of some diffeomorphism $f : \R^n \rightarrow \R^n$, equipped with coordinates $(x^a)$. It is constructed as follows,
\begin{equation}
\label{def_jet}
j^2_0(f) = \left(u^a + u^a_b x^b + \half u^a_{bc} x^b x^c\right) \in \R^n,
\end{equation}
where $u^a$, $u^a_b$ and $u^a_{bc} = u^a_{cb}$ are the coefficients of the Taylor expansion of $f$ at $x^a = 0$ up to order 2. Jets follow the known properties of Taylor expansions, in particular the composition $j^2_0(f) \circ j^2_0(g) = j^2_0(f \circ g)$. Thus, taking a second diffeomorphism $g$ that has a vanishing jet of zeroth order, \ie $j^2_0(g) = \left(s^a_b x^b + \half s^a_{bc} x^b x^c\right)$, we find for the composition,
\begin{equation}
\label{jet_action}
j^2_0(f \circ g) = \left(u^a + u^a_b s^b_c x^c + \half \left(u^a_b s^b_{cd} + u^a_{be} s^b_d s^e_c\right) x^d x^c \right) 
\end{equation}
The group $G^2(n)$ is then the set of $2$-jets at $0$ associated to all diffeomorphisms $f, f' : R^n \rightarrow \R^n$ with vanishing zeroth order, such that the group law is given by $j^2_0(f) \circ j^2_0(f') = j^2_0(f \circ f')$, for $j^2_0(f), j^2_0(f') \in G^2(n)$.

Now, as we have explained above, the 2-frame bundle $P^2(M) = P(M, G^2(n))$ is a principal $G^2(n)$-bundle. It is the set of $2$-jets at $0$ of all possible local diffeomorphisms $f : \R^n \rightarrow M$ such that $f(0) = x \in M$. The right action of $j^2_0(s) \in G^2(n)$ on $j^2_0(f) \in P^2(M)$ is given by $j^2_0(f) \circ j^2_0(s) = j^2_0(f \circ s)$.

In practice, instead of working with jets as denoted in \eqref{def_jet} we work directly with the coefficients defining the jets, \eg $(u^a, u^a_b, u^a_{bc})$. The natural coordinates of $P^2(M)$ are then $(x^\mu, {e^\mu}_a, e^\mu_{ab})$, where $(x^\mu)$ are the coordinates of the base $M$, and where indices denoted by greek letters refer to the base $M$, and latin indices refer to the fibers. $j^2_0(f)=(x^\mu, {e^\mu}_a, e^\mu_{ab})$ is said to be a $2$-frame at the point $x$. The right action of $a = (s^a_b, s^a_{bc}) \in G^2(n)$ on $(x^\mu, {e^\mu}_a, e^\mu_{ab}) \in P^2(M)$ is then $r_a (x^\mu, {e^\mu}_a, e^\mu_{ab}) = (x^\mu, {e^\mu}_a, e^\mu_{ab})(s^a_b, s^a_{bc}) = (x^\mu, {e^\mu}_c s^c_a, {e^\mu}_c s^c_{ab} + e^\mu_{cd} s^c_a s^d_b)$, and the group law is such that, for $(s^a_b, s^a_{bc}), (s'^a_b, s'^a_{bc}) \in G^2(n)$, we have $(s'^a_b, s'^a_{bc}) (s^a_b, s^a_{bc}) = (s'^a_d s^d_b, s'^a_d s^d_{bc} + s'^a_{de} s^d_b s^e_c)$. 

\medskip

On such bundles, one can define a canonical 1-form \cite{Kobayashi61} $(\theta^a, {\theta^a}_b) \in \Omega^1(P^2(M), \R^n\oplus\gl(n,\R))$, where we have in general, with coordinates such that $\theta^a = {\theta^a}_\mu dx^\mu$, and where ${\theta^a}_\mu$ is the inverse of ${e^\mu}_a$, \ie ${\theta^a}_\mu {e^\mu}_b = \delta^a_b$, and
\begin{equation}
\label{general_def_theta}
{\theta^a}_b = {\theta^a}_\lambda \left(d{e^\lambda}_b - e^\lambda_{bc} \theta^c\right).
\end{equation}
The first two components of the canonical form then obey the relation,
\begin{equation}
d\theta^a + {\theta^a}_b \wedge \theta^b = 0.
\end{equation}

In the following, we will consider sub-bundles of $P^2(M)$ to represent conformal and projective geometries\footnote{Note that in the following, we use notations where the subscript ${}_\cC$ means that the object refers to conformal geometry, and the subscript ${}_\cP$ will refer to projective geometry.}, together with the usual principal bundles where the structure group is in its fundamental matrix representation.

\subsection{Conformal geometry as a principal bundle... \label{subsect:conf-geom}}

The conformal geometry is based on the Klein model $(G_\cC, H_\cC)$ where is $G_\cC$ the M\"obius group and $H_\cC$ its homogeneous subgroup, \ie the subgroup leaving any point of $G_\cC/H_\cC$ invariant. 

\subsubsection{... using the usual matrix representation}

The M\"obius group $G_\cC = O(n, 2)/\{\pm I_{n+2}\} = O(n, 2)/ \mathbb{Z}_2$ is the group preserving the bilinear form
\begin{equation}
\Sigma = \left(\begin{matrix}
0 & 0 & -1 \\
0 & \eta & 0 \\
-1 & 0 & 0 
\end{matrix}\right),
\end{equation}
\ie $\forall a \in G_\cC, a^T \Sigma a = \Sigma$, with $^T$ denoting the transposition, and $\eta$ the flat metric of signature $(n-1,1)$, quotiented by the abelian group $\mathbb{Z}_2=\{\pm I_{n+2}\}$ \cite{Sharpe96}. This group is parametrized by isometries of the metric $\eta$, $S \in O(n-1,1)$, translations $t \in \R^{n-1,1}$, Weyl dilations $z \in W = \R\setminus \{0\}$ and special conformal transformations $r \in {\R^n}^*$. A representation of the group $G_\cC$ parametrized by these elements can be realized as a subgroup of $\GL(n+2)$ such that its elements are matrices of the form,
\begin{equation}
\label{def_G_matrix}
\left(\begin{matrix}
z & z r_b & z \frac{r^2}{2} \\
z t^a & {S^a}_b + z t^a r_b & z \frac{r^2}{2} t^a + {S^a}_c r^c \\
\frac{z t^2}{2} & \frac{z t^2}{2} r_b + t_c {S^c}_b & \frac{1}{z} + z \frac{t^2 r^2}{4} + t_c {S^c}_d r^d
\end{matrix}\right) = \left(\begin{matrix}
1 & 0 & 0 \\
t^a & \delta^a_c & 0 \\
\frac{t^2}{2} & t_c & 1
\end{matrix}\right)
\left(\begin{matrix}
z & 0 & 0 \\
0 & {S^c}_d & 0 \\
0 & 0 & \frac{1}{z}
\end{matrix}\right)
\left(\begin{matrix}
1 & r_b & \frac{r^2}{2} \\
0 & \delta^d_b & r^d \\
0 & 0 & 1
\end{matrix}\right)
\end{equation}
with $r^d := \eta^{db}r_b$, and $t_c := \eta_{ca}t^a$.

We see that an element in $G_\cC$ can be decomposed as a product of elements in $T$ the subgroup of translations, $\CO(n-1, 1) = O(n-1, 1) \times W$ the conformal group, with $W$ the Weyl group, and $K$ the subgroup of conformal special transformations.

The group law can be computed by standard matrix multiplication. If two group elements are written as $({S^a}_b, z, t^a, r_b)$ and $({S'^a}_b, z', t'^a, r'_b) \in G_\cC$, their product reads
\begin{equation}
\label{eq:conf-group-law}
\begin{array}{l}
\displaystyle ({S'^a}_b, z', t'^a, r'_b)\circ ({S^a}_b, z, t^a, r_b) =  \\
\displaystyle \qquad\qquad\qquad \left({S'^a}_c {S^c}_b, \frac{z' z(2+r'_c t^c)^2}{4} ,t'^a +  \frac{S'^a{}_c t^c}{z'(1+ \frac{r'_c t^c}{2})}, r_b + \frac{r'_c {S^c}_b}{z(1 + \frac{r'_c t^c}{2})} \right)
\end{array}
\end{equation}

Its subgroup $H_\cC$, which leaves invariant points of $G_\cC/H_\cC$, is the subgroup of $G_\cC$ without translations, and can be represented as \cite{Sharpe96},
\begin{equation}
\label{def_H_matrix}
H_\cC = \left\lbrace \left(\begin{matrix}
z & z r_b & z \frac{r^2}{2} \\
0 & {S^a}_b & {S^a}_c r^c \\
0 & 0 & z^{-1}
\end{matrix}\right) = 
\left(\begin{matrix}
z & 0 & 0 \\
0 & {S^c}_d & 0 \\
0 & 0 & z^{-1} 
\end{matrix}\right)
\left(\begin{matrix}
1 & r_b & \half r^2 \\
0 & \delta^d_b & r^d \\
0 & 0 & 1
\end{matrix}\right)
\right\rbrace.
\end{equation}
This subgroup has the decomposition,
\begin{equation}
\label{decomp_H}
H_\cC = \CO(n-1,1) \ltimes K = \left(O(n-1, 1) \times W\right) \ltimes K,
\end{equation}
whose elements can be written as $({S^a}_b, z, r_b):=({S^a}_b, z, t^a=0, r_b)$. From \eqref{eq:conf-group-law} the induced group law is given by
\begin{equation}
\label{group_law_H}
({S'^a}_b, z', r'_b) \circ ({S^a}_b, z, r_b ) =  \left({S'^a}_c {S^c}_b, z' z, r_b + z^{-1} r'_c {S^c}_b\right).
\end{equation}
Its well-known projective action on $x \in \R^{n-1,1}$ is then\footnote{One sometimes finds in the literature $r$ defined with the opposite sign.},
\begin{equation}
\label{action_mobius_rn}
x'^a = \frac{S^a{}_b x^b + S^a{}_b r^b \, x^2/2}{z(1+r_b x^b + r^2 x^2/4)}.
\end{equation}

One can define the Lie algebras associated to $G_\cC$ and $H_\cC$, which we denote respectively $\mathfrak{g}_\cC$ and $\mathfrak{h}_\cC$, by linearization of their matrix representation. We have then,
\begin{equation}
\label{def_lie_gh}
\mathfrak{g}_\cC = \left\lbrace \left(\begin{matrix}
{\alpha^0}_0 & {\alpha^0}_b & 0 \\
{\alpha^a}_0 & {\alpha^a}_b & \eta^{ac} {\alpha^0}_c \\
0 & \eta_{bc} {\alpha^c}_0 & - {\alpha^0}_0
\end{matrix}\right) \right\rbrace, \qquad
\mathfrak{h}_\cC = \left\lbrace \left(\begin{matrix}
{\alpha^0}_0 & {\alpha^0}_b & 0 \\
0 & {\alpha^a}_b & \eta^{ac} {\alpha^0}_c \\
0 & 0 & - {\alpha^0}_0
\end{matrix}\right) \right\rbrace, \Tr {\alpha^a}_b = 0.
\end{equation}

Both algebras are graded, \ie $\mathfrak{g} = \mathfrak{g}_{-1} \oplus \mathfrak{g}_0 \oplus \mathfrak{g}_1 = \R^n \oplus \mathfrak{co}(n-1,1) \oplus  {\R^n}^*$, and $\mathfrak{h} = \mathfrak{g}_0 \oplus \mathfrak{g}_1$, with $\left[g_i, g_j\right] \subset g_{i+j}$. This graduation neatly corresponds to the decomposition in terms of matrices in the definition \eqref{def_G_matrix}. We have indeed \cite{Kobayashi72},
\begin{equation}
\label{lie_graduation}
\begin{matrix}
\mathfrak{g_{-1}} = \left\lbrace \left(\begin{matrix}
0 & 0 & 0 \\
{\alpha^a}_0 & 0 & 0 \\
0 & \eta_{bc} {\alpha^c}_0 & 0
\end{matrix}\right) \right\rbrace, \quad \rule{0pt}{2em} & 
\mathfrak{g_{0}} = \left\lbrace \left(\begin{matrix}
{\alpha^0}_0 & 0 & 0 \\
0 & {\alpha^a}_b & 0 \\
0 & 0 & - {\alpha^0}_0
\end{matrix}\right) \right\rbrace, \, \\[5ex]
\mathfrak{g_{1}} = \left\lbrace \left(\begin{matrix}
0 & {\alpha^0}_b & 0 \\
0 & 0 & \eta^{ac} {\alpha^0}_c \\
0 & 0 & 0
\end{matrix}\right) \right\rbrace. \quad \rule{0pt}{2em} &
\end{matrix}
\end{equation}

\subsubsection{... using the 2-frames representation}

\medskip

Conformal 2-frame bundles $P_\cC$ are 2-frame bundles, as defined in section \ref{section_2-frames}, where we restrict the structure group, denoted $H^2_\cC(n)$, to be the subgroup of $G^2(n)$ such that it is isomorphic to the M\"obius group. One can define them intrinsically as in \cite{Ogiue67}, or using the link between the matrix and the 2-frame representations of the M\"obius group. Since we want to emphasis the link between the 2 representations in this paper, we will show the latter.

The conformal 2-frame bundle $P_\cC$ and the the principal bundle $P(M, G_\cC)$ of the previous section then describe the same geometry. Indeed, they have the same base $M$ and the same structure group, just with a different representation. The homomorphism between the two descriptions is obtained by computing the jet associated to the action of the group $H_\cC$ on the base, \ie the formula \eqref{action_mobius_rn}. The subgroup $H^2_\cC(n)$ of $G^2(n)$ is thus the realization of $H_\cC$ \eqref{def_H_matrix} in $G^2(n)$, described in \eqref{section_2-frames}, such that $(h^a_b, h^a_{bc}) \in H^2_\cC(n)$, with
\begin{subequations}
\label{hom_conf}
\begin{align}
{h^a}_b & = z^{-1} {S^a}_b \label{expr_1st_order}\\
h^a_{bc} & = z^{-1} \left( {S^a}_d \eta^{de} r_e \eta_{bc} - 2  r_{(b} {S^a}_{c)} \right) \label{expr_2nd_order}
\end{align}
\end{subequations}
for $(h^a_b, h^a_{bc}) \in H_\cC^2(n)$. Note that the first relation implies $\eta_{ab} {h^a}_c {h^b}_d = z^{-2} \eta_{cd}$, and the second one implies that $(h^a_{bc})$ only carries $n$ additional degrees of freedom. Indeed, one can write these coefficients as,
\begin{equation}
\label{rel_2nd_order}
h^a_{bc} := \eta^{de}\eta_{bc} h_d {h^a}_e - {h^a}_b h_c - {h^a}_c h_b
\end{equation}
for $h_a := r_a$. The above definition implies that elements in $H^2_\cC(n)$ may be instead parametrized as $(h^a_b, h_b)$. Natural coordinates on $P_\cC$ are then $(u^\alpha, u^\alpha_a, u_a)$. Note that if we write $(h^a_b, h_b) \in H^2_\cC(n)$, we have the group law, $(h'^a_b, h'_b)(h^a_b, h_b) = (h'^a_c h^c_b, h_b+h'_c h^c_b)$. This group law then matches group law of the subgroup $H_\cC$ of the M\"obius group \eqref{group_law_H}. We have then
\begin{equation}
\label{realization_h}
h = \left(z^{-1} {S^a}_b, r_b\right) \in H^2_\cC(n).
\end{equation}
This definition of the conformal 2-frame bundle $P_\cC$ matches with the intrinsic definition given in~\cite{Ogiue67}.

\smallskip
Consider now, at the level of Lie algebras, an element $\left(\widetilde{\alpha}^a_b, \widetilde{\alpha}^a_{bc}\right) \in \mathfrak{h^2}_\cC(n)$. Linearizing the homomorphism \eqref{hom_conf}, using the inverse relation
\begin{align}
\label{eq:paramG1}
h_b = - \frac{1}{n} h^a_{bc} h^c{}_a
\end{align}
and using the definitions \eqref{def_lie_gh} of $\mathfrak{g}_\cC$ and $\mathfrak{h}_\cC$, we find the relations,
\begin{subequations}
\label{alg_hom}
\begin{align}
{\walpha^a}_b & = {\alpha^a}_b - {\alpha^0}_0 \delta^a_b \label{alg_hom_1}\\
\walpha_b & = {\alpha^0}_b, \label{alg_hom_2}
\end{align}
\end{subequations}
and the linearized version of \eqref{rel_2nd_order} reads
\begin{equation}
\walpha^a_{bc} = \eta^{ad}\eta_{bc} \walpha_d - \delta^a_b \walpha_c - \delta^a_c \walpha_b\, .
\end{equation}

\medskip
From now on, we will take $M$ to be a Lorentzian spacetime with coordinates $(x^\mu)$. Coordinates on its associated conformal 2-frame bundle are then parametrized by $\left(x^\mu, {e^\mu}_a, e_a\right)$, and we have $\theta^a := {\theta^a}_\mu dx^\mu$ such that ${\theta^a}_\mu {e^\mu}_b = \delta^a_b$. Much like in \eqref{rel_2nd_order}, we can define coefficients $e_a$ such that
\begin{equation}
\label{rel_2nd_order_coords}
e^\mu_{ab} = \eta_{ab} \eta^{cd}{e^\mu}_c e_d - {e^\mu}_a e_b - {e^\mu}_b e_a,
\end{equation}
or equivalently $e_a = - \frac{1}{n} e^\mu_{ab} {\theta^b}_\mu$. 

By Gram-Schmidt, one can construct linear frames which are orthonormal with respect to a metric $g$ on $TM$, such that $g(e,e) = \eta$ at each point $x\in M$. In the local coordinates $(x^\mu)$ on $M$, one has 
\begin{equation}
\eta_{ab} {\theta^a}_\mu (x){\theta^b}_\nu (x)= g_{\mu\nu} (x)= g({\partial_\mu}_{|x},{\partial_\nu}_{|x})
\label{realisation_metric}
\end{equation}
or equivalently, $g_{\mu\nu}(x) {e^\mu}_a(x) {e^\nu}_b(x) = \eta_{ab}$. Note that in these two relations between $g$ and $\eta$ the vierbein ${e^\mu}_a(x)$ can be seen as part of the section $s: (x^\mu) \mapsto (x^\mu, {e^\mu}_a(x), e_a(x))$ of $P_{\cal C}$, hence the dependence on the base coordinates $(x^\mu)$. In the rest of the paper, ${e^\mu}_a$ will denote the coordinate on the conformal 2-frame bundle while ${e^\mu}_a(x)$ will denote the element of the section.

\subsubsection{Connection on a conformal 2-frame bundle}

We have seen two ways of defining a conformal structure, either \textit{via} a principal bundle $P(M, H_\cC)$, or \textit{via} a principal bundle $P(M, H^2) = P_\cC$, the later which we called conformal 2-frame bundle. We have also seen the homomorphism to go from one description to the other. Accordingly, we can define a normal Cartan connection both using the first description, in terms of matrices, and using the second description in terms of frames\footnote{We will use notations such that tilde variables are variables for the 2-frame representation, whereas non tilde variables are variables for the matrix representation.}.

\smallskip
Let us recall the definition of a Cartan connection on a fiber bundle $P(M, H_\cC)$, based on the Klein geometry $(G_\cC,H_\cC)$: it is a 1-form $\omega \in \Omega^1(P(M, H_\cC), \mathfrak{g}_\cC)$, such that,\begin{subequations}
\label{cartan_connection}
\begin{align}
r^*_h \omega & = \Ad(h^{-1}) \omega, & &\forall \, h \in H_\cC \label{equiv_cartan} &\\
\omega(X_A) & = A, & &\forall \, A \in \mathfrak{h}_\cC &\\
\omega(X) & \neq 0, & &\forall \, X \neq 0,\ X\in \Gamma(TP) &
\end{align}
\end{subequations}
where $X_A$ is the fundamental vertical vector field generated by the Lie algebra element $A$. The curvature form $\Omega\in \Omega^2(P(M, H_\cC), \mathfrak{g}_\cC)$ of the Cartan connection $\omega$ is then defined \textit{via} the structure equations,
\begin{equation}
d\omega = - \half [\omega, \omega] + \Omega.
\end{equation}

In the matrix representation, a connection $\omega \in \Omega^1(P(M, H_\cC), \lieg_\cC)$ can be parametrized in full generality as,
\begin{equation}
\label{matrix_rep}
\omega = \left(\begin{array}{ccc}
{\omega^0}_0 & {\omega^0}_b & 0 \\
{\omega^a}_0 & {\omega^a}_b & \eta^{ac} {\omega^0}_c \\
0 & \eta_{cb} {\omega^c}_0 & - {\omega^0}_0
\end{array}\right), \quad  \Tr {\omega^a}_b = 0
\end{equation}

On the 2-frame bundle a connection is given by $\womega = \left(\womega^a, {\womega^a}_b, \womega_b\right) \in \Omega^1(P(M, H^2), \lieg^2_\cC(n))$, where the algebra is simply given by $\lieg^2_\cC(n) = \R^n \oplus \mathfrak{h}^2_\cC(n)$. Since we have an homomorphism between the two descriptions, see \eqref{alg_hom}, we can relate the coefficients of the first connection \eqref{matrix_rep} with those of the second connection, with, 
\begin{subequations}
\label{hom_conf_connection}
\begin{align}
\womega^a & = {\omega^a}_0, \label{hom_debut} \\
{\womega^a}_b & = {\omega^a}_b - \delta^a_b {\omega^0}_0, \\
\womega_b & = {\omega^0}_b. \label{hom_fin}
\end{align}
\end{subequations}

Since $({\omega^a}_b)$ is traceless by definition, it is possible to inverse the above relations. We have indeed,
\begin{subequations}
\label{conf_inv_rel}
\begin{align}
{\omega^a}_0 & = \womega^a \label{inv_rel_debut}\\
{\omega^0}_0 & = - \frac{1}{n} \delta^b_a {\womega^a}_b \\
{\omega^a}_b & = {\womega^a}_b + \delta^a_b {\omega^0}_0 \\
{\omega^0}_b & = \womega_b \, . \label{inv_rel_fin}
\end{align}
\end{subequations}

\medskip

Following the ``Japanese school'' on connections on 2-frame bundles \cite{Kobayashi61}, in order to find the normal connection on a conformal 2-frame bundle $P_\cC$, and according to \cite[appendix I]{TheseChristian}, one sets 
\begin{subequations}
\begin{align}
\womega^a & = \theta^a  \\
{\womega^a}_b & = {\theta^a}_b + {\varphi^a}_b \label{def_womegaab}
\end{align}
\end{subequations}
for the canonical form $(\theta^a, {\theta^a}_b) \in \Omega^1(P_\cC, \R^n \oplus \mathfrak{co}(n))$, and for some ${\varphi^a}_b$. Given \eqref{general_def_theta} and \eqref{rel_2nd_order_coords}, we have,
\begin{equation}
\label{def_thetaab}
{\theta^a}_b = {\theta^a}_\mu d{e^\mu}_b - \eta^{ac}\eta_{bd}e_c \theta^d + e_b\theta^a + e_c \theta^c \delta^a_b\,.
\end{equation}
One then finds the corresponding unique $\womega_b$ such that the connection is normal. Recall that a normal Cartan connection $\womega$ is a Cartan connection such that its curvature $\wOmega = (\wOmega^a, {\wOmega^a}_b, \wOmega_b) \in \Omega^2(P, \lieg^2_\cC(n))$ has vanishing torsion, \ie $\wOmega^a = 0$, and satisfies ${K^a}_{bac} = 0$, where  ${\wOmega^a}_b = \frac{1}{2}{K^a}_{bcd} \, \womega^c \wedge \womega^d$ \cite{Kobayashi72}. This normal Cartan connection is then unique. In the end, one has \cite{Ogiue67}, (see also Appendix~\ref{sect:appendix} for a detailed account on how to get such local expressions on the conformal 2-frame bundle)
\begin{subequations}
\label{eq:normal-Cartan}
\begin{align}
\womega^a & = \theta^a \label{def_wa} \\
{\womega^a}_b & = {\theta^a}_\mu d{e^\mu}_b - \eta^{ac}\eta_{bd} e_c \theta^d + e_b \theta^a + \left(e_c \theta^c\right)\delta^a_b + {\theta^a}_\mu \Pi^\mu_{\nu\lambda} {e^\nu}_b dx^\lambda \label{def_wab} \\
\womega_b & = de_b - e_c \womega^c_b + e_c \theta^c e_b + {e^\mu}_b \Pi_{\mu\nu} dx^\nu - \half \eta^{cd}\eta_{be} e_c e_d \theta^e, \label{def_wb}
\end{align}
\end{subequations}
with ${\varphi^a}_b := {\theta^a}_\mu \Pi^\mu_{\nu\lambda} {e^\nu}_b dx^\lambda$, and where $\Pi^\mu_{\nu\lambda}$ and $\Pi_{\mu\nu}$ are \textit{a priori} functions on the conformal 2-frame bundle $P_\cC$.

\smallskip
Let us now prove that $\Pi^\mu_{\nu\lambda}$ and $\Pi_{\mu\nu}$ are in fact functions of the base only. Since $\omega$ is a Cartan connection, we have by definition the right action of $H$ \eqref{def_H_matrix} on the connection, with $h \in H_\cC$,
\begin{equation}
\label{right_action_connection}
r^*_h \omega = \Ad(h^{-1}) \omega.
\end{equation}
Given an element $h \in H_\cC$ \eqref{def_H_matrix}, the matrix representation of the connection $\omega$ \eqref{matrix_rep}, and the homomorphism \eqref{hom_conf_connection}, we find the adjoint action on the representation $\womega$ of the connection,
\begin{subequations}
\begin{align}
\Ad(h^{-1}) \womega^a & = z {(S^{-1})^a}_c \womega^c, \label{ad_wa} \\
\Ad(h^{-1}) {\womega^a}_b & = {(S^{-1})^a}_c \left( {\womega^c}_d {S^d}_b + z \womega^c r_b \right) - z \eta^{ac} r_c \womega^d \eta_{de} {S^e}_b + z \delta^a_b r_c {(S^{-1})^c}_d \womega^d, \label{ad_wab} \\
\Ad(h^{-1}) \womega_b & = - r_c {(S^{-1})^c}_d \left({\womega^d}_e {S^e}_b + \womega^d r_b\right) + z^{-1} \womega_d {S^d}_b + \frac{z}{2} r^d \eta_{de} r^e \womega^a \eta_{ac} {S^c}_b. \label{ad_wb}
\end{align}
\end{subequations}

Now, as we have seen in section \ref{section_2-frames}, the right action of $h \in H_\cC$ \eqref{realization_h} on the coordinates of the conformal 2-frame bundle is
\begin{equation}
\label{right_action_h_conf}
r_h \left(x^\mu, {e^\mu}_a, e_b\right) = \left(x^\mu, z^{-1} {e^\mu}_c {S^c}_a, r_b + z^{-1} e_c {S^c}_b\right).
\end{equation}
Of particular interest is the action on ${e^\mu}_a$, which when applied to the identity ${e^\mu}_c {\theta^c}_\nu = \delta^\mu_\nu$ gives the action on $\theta^a{}_\nu$,
\begin{equation}
r_h {\theta^a}_\nu = z{(S^{-1})^a}_c {\theta^c}_\nu.
\end{equation}
We then compute the right action of $h$ on the canonical form $(\theta^a, {\theta^a}_b)$. We find,
\begin{subequations}
\begin{align}
r_h \theta^a & = z{(S^{-1})^a}_c {\theta^c}, \\
r_h {\theta^a}_b & = {(S^{-1})^a}_c {\theta^c}_d {S^d}_b + z r_b {(S^{-1})^a}_c \theta^c - z \eta^{ac} r_c  \eta_{bd} {(S^{-1})^d}_e \womega^e + z \delta^a_b r_c {(S^{-1})^c}_d \theta^d. \label{rhthetaab}
\end{align}
\end{subequations}
Keeping in mind the defining relation of $S \in O(n-1,1)$, we find the same form for the right action $r_h {\theta^a}_b$ \eqref{rhthetaab} and the adjoint action $\Ad(h^{-1}) {\womega^a}_b$, see \eqref{ad_wab}. Then, given our definition \eqref{def_womegaab} of ${\womega^a}_b$, the relation \eqref{right_action_connection} leads to $r_h {\varphi^a}_b = {(S^{-1})^a}_c {\varphi^c}_d {S^d}_b$. Hence, if ${\varphi^a}_b = {\theta^a}_\mu \Pi^\mu_{\nu\lambda} {e^\nu}_b dx^\lambda$, we find,
\begin{equation}
r_h \Pi^\mu_{\nu\lambda} = \Pi^\mu_{\nu\lambda}.
\end{equation}
Given \eqref{right_action_h_conf}, this means the $\Pi^\mu_{\nu\lambda}$ are functions of the base only, \ie $\Pi^\mu_{\nu\lambda} = \Pi^\mu_{\nu\lambda}(x)$.

Similarly for the second order part of $\womega$, namely $\womega_b$, equating the adjoint action \eqref{ad_wb} and the right action of $h \in H_\cC$ on  \eqref{def_wb}, leads to 
\begin{equation}
r_h \Pi_{\mu\nu} = \Pi_{\mu\nu},
\end{equation}
meaning, again, that the $\Pi_{\mu\nu}$ are functions of the base only, \ie $\Pi_{\mu\nu} = \Pi_{\mu\nu}(x)$. 

\medskip 
From the expressions of the elements of the connection in the 2-frame representation \eqref{eq:normal-Cartan} and the homomorphism \eqref{conf_inv_rel}, we find the expression of the parameters of the connection in the matrix representation \eqref{matrix_rep},
\begin{subequations}
\label{conf_connection}
\begin{align}
{\omega^a}_0 & = \theta^a, \label{wa0} \\
{\omega^0}_0 & = - e_a \theta^a - \frac{1}{n} \left({\theta^a}_\mu d{e^\mu}_a + \Pi^\mu_{\mu\lambda} dx^\lambda\right) := -e_a \theta^a - T, \label{w00} \\
{\omega^a}_b & = {\theta^a}_\mu d{e^\mu}_b - \eta^{ac}\eta_{bd} e_c \theta^d + e_b \theta^a + {\theta^a}_\mu \Pi^\mu_{\nu\lambda} {e^\nu}_b dx^\lambda - T \delta^a_b, \label{wab} \\
{\omega^0}_b & = de_b - e_c {\theta^c}_\mu d{e^\mu}_b -e_c\theta^c e_b + \half e_c \eta^{cd}e_d \eta_{ba} \theta^a - e_c {\theta^c}_\mu \Pi^\mu_{\nu\lambda} {e^\nu}_b dx^\lambda + {e^\mu}_b \Pi_{\mu\lambda} dx^\lambda, \label{w0b}
\end{align}
\end{subequations}
for some functions $\Pi^\mu_{\nu\lambda}$ and $\Pi_{\mu\lambda}$ on $M$.

\subsection{Projective geometry}

For the projective geometry, one thus mimics the account given in Section \ref{subsect:conf-geom} for the conformal geometry.

\subsubsection{Matrix representation}

Projective structures are based on the Klein model $(G_\cP, H_\cP)$ with $G_\cP = \PSL(n, \R)$ and $H_\cP$ its homogeneous subgroup. Some authors work with $\PSL(n, \R)$, \eg \cite{Sharpe96, TheseCarina}, while some work with $\PGL(n, \R)$, \eg \cite{KobayashiN64,Kobayashi72}. These two groups do not differ much, however, as $\PSL(n, \R)$ is the identity component of $\PGL(n, \R)$, and the later is connected if $n$ is even and has two components if $n$ is odd \cite{Sharpe96}. We will use here $G_\cP = \PSL(n, \R)$ for convenience. The projective special linear group is isomorphic to \cite{Sharpe96},
\begin{equation}
\label{def_projective_g}
G_\cP = \left\lbrace a = \left(\begin{matrix}
{S^a}_b & c^a \\
b_b & d
\end{matrix}\right) \mod Z , \text{ with } \det a = 1\right\rbrace,
\end{equation}
where $Z$ is the center of $\SL(n+1, \R)$, \ie $Z = I_{n+1}$ if $n$ is even and $Z = \{\pm I_{n+1}\}$ if $n$ is odd. We have $S \in \GL(n)$, $b, c \in \R^n$, and $d \in \R \setminus \{0\}$.

This group acts projectively on the projective space $P^n(\R) \simeq G_\cP/H_\cP$. If $y \in \R^n$, we have,
\begin{equation}
\label{proj_action}
y'^a = \frac{{S^a}_b y^b + c^a}{b_b y^b + d},
\end{equation}
and the group law is given by matrix multiplication.

The group $H_\cP$ is the subgroup of $G_\cP$ leaving points of $G_\cP/H_\cP$ invariant. From the action \eqref{proj_action}, we see that this corresponds to the subgroup defined by $c^a = 0$ (no translations) and which leaves invariant the origin $0$. We will later use the decomposition as (semi) direct products of this group,
\begin{equation}
\label{proj_decomp_h}
H_\cP = (\GL(n,\R) \times \R \setminus \{0\}) \ltimes {\R^n}^*.
\end{equation}

The algebra of the projective group, much like the one of the conformal group, is graded. We have indeed, $\mathfrak{g}_\cP = \mathfrak{g}_{-1} \oplus \mathfrak{g}_0 \oplus \mathfrak{g}_1$, with,
\begin{equation}
\mathfrak{g}_{-1} = \left\lbrace \left(\begin{matrix}
0 & {\alpha^a}_0 \\
0 & 0
\end{matrix}\right)\right\rbrace,
\quad
\mathfrak{g}_{0} = \left\lbrace \left.a = \left(\begin{matrix}
{\alpha^a}_b & 0 \\
0 & {\alpha^0}_0
\end{matrix}\right)\right| \Tr a = 0\right\rbrace,
\quad
\mathfrak{g}_1 = \left\lbrace \left(\begin{matrix}
0 & 0 \\
{\alpha^0}_b & 0
\end{matrix}\right)\right\rbrace,
\end{equation}
and of course $\mathrm{Lie} \ H_\cP := \mathfrak{h}_\cP = \mathfrak{g}_0 \oplus \mathfrak{g}_1$.

Any 1-form $\alpha \in \Omega^1(P(M, H_\cP), \mathfrak{g}_\cP)$ on the principal fiber bundle $P(M, H_\cP)$ can hence be written in the representation given by this decomposition. To give an example, let us define the Cartan connection $\omega \in \Omega^1(P(M, H_\cP), \mathfrak{g}_\cP)$ in full generality as,
\begin{equation}
\label{proj_matrix_rep}
\omega = \left(\begin{array}{cc}
{\omega^a}_b & {\omega^a}_0 \\
{\omega^b}_0 & {\omega^0}_0
\end{array}\right),
\end{equation}
with ${\omega^0}_0 = - \delta^b_a {\omega^a}_b$ (the trace).

\subsubsection{2-frames representation}
\label{s:proj_connection}

Similarly to the conformal geometry case, one can also use the framework of the 2-frame bundle to describe the projective geometry. Projective 2-frame bundles $P_\cP$ \cite{KobayashiN64} are 2-frame bundles, as in Section~\ref{section_2-frames}, which model the jets associated to the projective group \eqref{def_projective_g} \cite{TheseCarina}. The representation of the projective group as a subgroup of the 2-frame group $G^2(n)$ is obtained by computing the jets associated to the action of the group $H$ \eqref{proj_decomp_h} on the base, \ie \eqref{proj_action} with $c^a = 0$. We have, for $(h^a_b, h^a_{bc}) \in H_\cP^2(n)$, and with the parametrization \eqref{def_projective_g},
\begin{subequations}
\begin{align}
{h^a}_b & = d^{-1} {S^a}_b, \\
h^a_{bc} & = - 2 d^{-2} b_{(b} {S^a}_{c)}\,.
\end{align}
\end{subequations}
Once again, elements in $H_\cP^2(n)$ can instead be parametrized as $(h^a_b, h_b) \in H_\cP^2(n)$, such that here $h^a_{bc} := - {h^a}_b h_c - {h^a}_c h_b$. Coordinates on projective 2-frame bundles are thus parametrized by $(x^\mu, {e^\mu}_a, e^\mu_{ab})$, or $(x^\mu, {e^\mu}_a, e_a) \in P_\cP = P(M, H^2_\cP(n))$, where $e^\mu_{ab} = -{e^\mu}_a e_b - {e^\mu}_b e_a$, or conversely $e_a = - \frac{1}{n+1} e^\mu_{ab} {e^b}_\mu$.

\medskip
And yet, we have a connection on the projective 2-frame bundle, denoted $\womega = (\womega^a, {\womega^a}_b, \womega_b)$ with an homomorphism between $\womega$ and $\omega$ which, in our notations, has the same formal expression as for the conformal case,
\begin{subequations}
\label{proj_hom}
\begin{align}
\womega^a & = {\omega^a}_0, \label{proj_hom_debut} \\
{\womega^a}_b & = {\omega^a}_b - \delta^a_b {\omega^0}_0, \\
\womega_b & = {\omega^0}_b. \label{proj_hom_fin}
\end{align}
\end{subequations}

The connection has almost the same expression as for the conformal case, see \eqref{eq:normal-Cartan}. Indeed, we have $\womega = (\theta^a, {\theta^a}_b + {\varphi^a}_b, \womega_b)$, with $(\theta^a, {\theta^a}_b)$ the canonical form, and thus \cite{KobayashiN64},
\begin{subequations}
\label{proj_def_connection}
\begin{align}
\womega^a & = \theta^a \label{proj_def_wa} \\
{\womega^a}_b & = {\theta^a}_\mu d{e^\mu}_b + e_b \theta^a + \left(e_c \theta^c\right)\delta^a_b + {\theta^a}_\mu \Pi^\mu_{\nu\lambda} {e^\nu}_b dx^\lambda \label{proj_def_wab} \\
\womega_b & = de_b - e_c {\womega^c}_b + e_c \theta^c e_b + {e^\mu}_b \Pi_{\mu\nu} dx^\nu, \label{proj_def_wb}
\end{align}
\end{subequations}
where ${\varphi^a}_b := {\theta^a}_\mu \Pi^\mu_{\nu\lambda} {e^\nu}_b dx^\lambda$. Using the homomorphism \eqref{proj_hom} we have for the components of the connection in the matrix representation \eqref{proj_matrix_rep},
\begin{subequations}
\label{proj_connection}
\begin{align}
{\omega^a}_0 & = \theta^a, \label{proj_wa0} \\
{\omega^0}_0 & = - e_a \theta^a - \frac{1}{n+1} \left({\theta^a}_\mu d{e^\mu}_a + \Pi^\mu_{\mu\lambda} dx^\lambda\right) := -e_a \theta^a - T, \label{proj_w00} \\
{\omega^a}_b & = {\theta^a}_\mu d{e^\mu}_b + e_b \theta^a + {\theta^a}_\mu \Pi^\mu_{\nu\lambda} {e^\nu}_b dx^\lambda - T \delta^a_b, \label{proj_wab} \\
{\omega^0}_b & = de_b - e_c {\theta^c}_\mu d{e^\mu}_b -e_c\theta^c e_b - e_c {\theta^c}_\mu \Pi^\mu_{\nu\lambda} {e^\nu}_b dx^\lambda + {e^\mu}_b \Pi_{\mu\lambda} dx^\lambda, \label{proj_w0b}
\end{align}
\end{subequations}
For some coefficients $\Pi^\mu_{\nu\lambda}$ and $\Pi_{\mu\lambda}$, which could be shown to be, once again, functions of the base $M$ only.

\section{The dressing field method}
\label{section_dressing}

In this section, we will apply the \emph{dressing field method} \cite{TheseJordan}, that we are quickly going to review, to reduce the gauge symmetries of the conformal structure, and then of the projective structure. We will then define conformal and projective tractors. 

\subsection{Review of the dressing field method}
\label{ss:review_dressing}

Since the dressing field method \cite{TheseJordan} is a quite recent process, we are going to quickly review it here, by applying it to the well-known Lorentzian geometry of General Relativity. The goal of this process is to reduce the gauge symmetries present in objects defined on the structure, such as the Cartan connection or the Cartan curvature, as one usually does physics with gauge invariant quantities. This section is merely a summary of \cite[chap.~2]{TheseJordan}, see the linked reference for more details, {\em e.g.} \cite{FournelFLM14}. 

\medskip
The Cartan geometry here is based on the Klein pair $(G,H)$ where $G = \SO(n-1, 1) \ltimes R^{n-1,1}$ is the Poincaré group and $H = \SO(n-1, 1)$ the Lorentz group, such that the homogeneous space $G/H = \R^{n-1,1}$ is the Minkowski spacetime. We then construct the principal fiber bundle $P(M, \SO(n-1,1))$, and equip it with a Cartan connection $\omega \in \Omega^1(P(M, \SO), \lieg)$, as defined in \eqref{cartan_connection}, where $\lieg$ is the Poincaré Lie algebra. Such a Cartan connection is written in full generality as,
\begin{equation}
\label{connection_poincare}
\omega = \left(\begin{matrix}
{\bar{\omega}^a}_b & \bar{\theta}^a \\
0 & 0
\end{matrix}\right),
\end{equation}
where $({\bar{\omega}^a}_b) \in \Omega^1(P(M, \SO), \so)$ and $(\bar{\theta}^a) \in \Omega^1(P(M, \SO), \R^{n-1,1})$ is the soldering form. 

Now, to apply the dressing field method, one needs to work locally. Hence, we need to pull-back the connection onto the base manifold with the help of a section $s : U \subset M \rightarrow P(M, SO)$. We write,
\begin{equation}
s^* \omega := \varpi = \left(\begin{matrix}
{A^a}_b & \theta^a \\
0 & 0
\end{matrix}\right),
\end{equation}
where $({A^a}_b)$ is the well-known usual spin connection and $(\theta^a)$ the vierbein 1-form. The problem with the above (local) connection is that it has a non trivial transformation under Lorentz gauge transformations $\gamma := \left(\begin{matrix}
{S^a}_b & 0 \\
0 & 1
\end{matrix}\right) \in \mathcal{SO}$. Indeed, one has $\varpi^\gamma = \gamma^{-1} \varpi \gamma + \gamma^{-1} d\gamma$, and so,
\begin{equation}
\label{trsf_lorentz}
\varpi^\gamma = \left(\begin{matrix}
{(S^{-1})^a}_c {A^c}_d {S^d}_b + {(S^{-1})^a}_c d{S^c}_b & {(S^{-1})^a}_c \theta^c \\
0 & 0
\end{matrix}\right).
\end{equation}

The idea of the dressing field method is to construct a (local) object that is gauge invariant, with the help of a \emph{dressing field} $u : U \subset M \rightarrow SO(n-1, 1)$, which has the defining equivariance under gauge transformations $u^\gamma := \gamma^{-1} u$. We then define the \emph{dressed connection} $\varpi_0 := u^{-1} \varpi u + u^{-1} du$. While this looks like a gauge transformation, this is not one, as $u$ is not in the gauge group. The resulting object is now indeed invariant under $\gamma$-gauge transformations, as ${\varpi}^{\gamma} = (u^{\gamma})^{-1} \varpi^{\gamma} u^{\gamma} + (u^{\gamma})^{-1} d(u^{\gamma}) = (\gamma^{-1} u)^{-1} \left(\gamma^{-1} \varpi \gamma + \gamma^{-1} d \gamma\right) \gamma^{-1} u + (\gamma^{-1} u)^{-1} d(\gamma^{-1} u) = u^{-1} \varpi u + u^{-1} du = \varpi$.

The dressing field is built from objects already present in the geometry, typically appearing in the Cartan connection itself. In this case, the (co)-vierbein $({\theta^a}_\mu)$ from the 1-form $\theta^a = {\theta^a}_\mu dx^\mu$ is used, where $(x^\mu)$ are coordinates on $M$. The dressing field $u$ is then
\begin{equation}
u = \left(\begin{matrix}
{\theta^a}_\mu & 0 \\
0 & 1
\end{matrix}\right).
\end{equation}
Indeed, we see from \eqref{trsf_lorentz} that this dressing field has the right equivariance: $u^\gamma = \gamma^{-1} u$. It is then possible to compute explicitly the dressed connection, which leads to, with ${e^\mu}_a$ the vierbein,
\begin{equation}
\varpi_0 = \left(\begin{matrix}
{e^\mu}_a {A^a}_b {\theta^b}_\nu + {e^\mu}_c \partial_\lambda {\theta^c}_\nu dx^\lambda & dx^\mu \\
0 & 0
\end{matrix}\right) := 
\left(\begin{matrix}
\Gamma^\mu_{\nu\lambda} & \delta^\mu_\lambda \\
0 & 0
\end{matrix}\right) dx^\lambda,
\end{equation}
where $\Gamma^\mu_{\nu\lambda}$ are the well-known Christoffel symbols. It is also possible to apply a very similar process to the Cartan curvature, leading to the definitions of the Riemann curvature tensor and the torsion tensor.

\medskip

In some situations, like in this paper, it is convenient to separate the dressing process into consecutive dressings. Indeed, take a geometry where, for instance, the structure group is given by $H = K_0 \ltimes K_1$. It will then typically be more convenient to first reduce the gauge symmetries associated to $K_1$, then those associated to $K_0$, rather than reduce everything at once. In practice, one could consider a dressing field $u : U \subset M \rightarrow H$, and define the dressed connection as $\varpi_0 := u^{-1} \varpi u + u^{-1} du$. Rather, one chooses to factor the dressing field as $u = u_0 u_1$, where $u_i : U \subset M \rightarrow K_i, \, i =0, 1$. We have then the partially dressed connection $\varpi_1 := u_1^{-1} \varpi u_1 + u_1^{-1} du_1$, and recover $\varpi_0 = u_0^{-1} \varpi_1 u_0 + u_0^{-1} du_0$, provided $u_1$ and $u_0$ fulfill some \emph{compatibility conditions}. 

First, both $u_i$ have to have the right equivariance under their own associated gauge group $\mathcal{K}_i$, \ie $u_i^{\gamma_i} = \gamma_i^{-1} u_i, \, i =0, 1$. Then, the first dressing field needs to transform as $u_1^{\gamma_0} = \gamma_0^{-1} u_1 \gamma_0$ under the second gauge transformations, and the second dressing field needs to be invariant, \ie $u_0^{\gamma_1} = u_0$, under the first gauge transformations. These requirements ensure that the connection $\varpi_1$ has a gauge-like (residual) transformation under $\mathcal{K}_0$, \ie $\varpi_1^{\gamma_0} := \gamma_0^{-1} \varpi_1 \gamma_0 + \gamma_0^{-1} d\gamma_0$, and that the fully dressed connection $\varpi_0$ is invariant under both $\mathcal{K}_0$ and $\mathcal{K}_1$. This process can be generalized to a split in an arbitrary number of steps, as shown in \cite{TheseJordan}.

\subsection{Conformal case}
\label{ss:conf}
\subsubsection{Christoffel symbols, Schouten tensor, and pull-back}

The coefficients $\Pi^\lambda_{\mu\nu}$ and $\Pi_{\mu\nu}$ appearing in this expression of the connection \eqref{conf_connection} are not at this stage, respectively, the Christoffel symbols or the Schouten tensor, but they are related to these objects by a conformal transformation. Indeed, they are such that the $\Pi^\lambda_{\mu\nu}$ are traceless, \textit{i.e.}, $\Pi^\rho_{\rho\mu} = 0$ \cite{Ogiue67}. The transformation is thus, 
\begin{equation}
\label{link_pi_christo}
\Pi^\rho_{\mu\nu} = \Gamma^\rho_{\mu\nu} - \frac{1}{n} \left(\delta^\rho_\mu \Gamma^\lambda_{\lambda\nu} + \delta^\rho_\nu \Gamma^\lambda_{\lambda\mu} - g^{\rho\kappa}\Gamma^\lambda_{\lambda\kappa} g_{\mu\nu}\right).
\end{equation}
Let us remark that the latter measures the default for an \emph{arbitrary} $2$-frame section 
\[
x \mapsto (x^\mu, e^\mu{}_a(x),  e^\mu_{ab}(x))
\]
to be a conformal frame, see \eqref{rel_2nd_order_coords}. Indeed, if one sets 
$\Gamma^\rho_{\mu\nu} (x) := - e^\rho_{ab}(x)\theta^a{}_\mu(x) \theta^b{}_\nu(x)$, then (dropping the $x$-dependence) $e^\rho_{ra}\theta^r{}_\rho = - e^\mu{}_a \Gamma^\lambda_{\lambda\mu}$\footnote{Notice also the relationships $\Gamma^\lambda_{\lambda\mu} = n e_a \theta^a{}_\mu$ and $\Gamma^\rho_{\mu\nu} = e^\rho{}_a \theta^a{}_{\mu\nu}$, the latter yields the well-known gluing rules for the $\Gamma$'s under a change of local coordinates on the base manifold $M$.}, and the difference
\[
e^\mu_{ab} - \frac{1}{n} \big( 
e^\mu{}_a (e^\rho_{rb}\theta^r{}_\rho) + e^\mu{}_b (e^\rho_{ra}\theta^r{}_\rho)  - \eta_{ab}\eta^{cd} e^\mu{}_c (e^\rho_{rd}\theta^r{}_\rho) \big)
\]
becomes exactly (up to a global minus sign) the above equation~\eqref{link_pi_christo} upon using the relation~\eqref{realisation_metric} for the metric. Moreover, the expression~\eqref{link_pi_christo} turns out to correspond to a rescaling of the metric,
\begin{equation}
\label{link_metrics}
\widetilde{g} = \Omega^2 g,
\end{equation}
with $\Omega = \vert g \vert^{-1/2n}$, and where $\Pi$ and $\Gamma$ are, respectively, the Christoffel symbols associated to the metrics $\widetilde{g}$ and $g$, in other words, the coefficients of the corresponding Levi-Civita connections. It is worthwhile to notice that the geometric object $\widetilde{g}$ defined in \eqref{link_metrics} is invariant under Weyl rescalings $g\to \lambda^2 g$. We will hence use the shorthand $\Upsilon_\mu := \Omega^{-1} \partial_\mu \Omega = - \frac{1}{n} \partial_\mu \ln \sqrt{\vert g\vert} = - \frac{1}{n} \Gamma^\lambda_{\lambda\mu}$, such that
\[
\Pi^\rho_{\mu\nu} = \Gamma^\rho_{\mu\nu} + \delta^\rho_\mu \Upsilon_\nu + \delta^\rho_\nu \Upsilon_\mu - g^{\rho\lambda}\Upsilon_\lambda g_{\mu\nu}\ .
\]

In the same spirit, $\Pi_{\mu\nu}$ transforms as the Schouten tensor under the transformation \eqref{link_metrics},
\begin{equation}
\label{link_pi_schouten}
\Pi_{\mu\nu} = P_{\mu\nu} + \nabla_\mu \Upsilon_\nu - \Upsilon_\mu \Upsilon_\nu + \half g^{\lambda\rho} \Upsilon_\lambda \Upsilon_\rho g_{\mu\nu},
\end{equation}
where $P = \frac{-1}{n-2}\left(\Ric - \frac{R}{2(n-1)}g\right)$ will be the Schouten tensor related to the metric $g$.

\smallskip
The 1-form $T$ defined through \eqref{w00} hence simplifies to,
\begin{equation}
\label{simp_T}
T = \frac{1}{n} {\theta^c}_\rho d{e^\rho}_c.
\end{equation}

Using the expression of the metric \eqref{realisation_metric} in terms of the 2-frame bundle elements, we find the local expression for the trace of the Christoffel symbols associated to the metric $g$,
\begin{equation}
\label{local_gamma}
\Gamma^\rho_{\rho\mu}(x) = - {\theta^c}_\rho(x) \partial_\mu {e^\rho}_c(x).
\end{equation}

Now, the dressing field method is a local process, so we will need a section $s$ to pull back the section onto the base, just like when pullbacking the metric onto the base, see \eqref{realisation_metric}. We will write $s^* \omega \rightarrow \omega$, and drop the dependency in the coordinates of $e_b$ and ${e^\mu}_a$, $e_b(x) \rightarrow e_b$ and  ${e^\mu}_a(x) \rightarrow {e^\mu}_a$. In the following, $d$ will now be the differential of $M$.

From \eqref{simp_T} and \eqref{local_gamma}, we find that $s^* T_\mu = - \frac{1}{n} \Gamma^\rho_{\rho\mu}$, and thus we can relate $t$ to the previously defined $\Upsilon$,
\begin{equation}
\label{upsilon_t}
\Upsilon = s^* T.
\end{equation}

Using the expressions \eqref{link_pi_christo}, \eqref{link_pi_schouten} and \eqref{upsilon_t}, we can rewrite the local expression of the connection \eqref{conf_connection} as $\varpi := s^* \omega$, such that,
\begin{subequations}
\label{local_conf_connection}
\begin{align}
{\varpi^a}_0 & := s^* {\omega^a}_0 = \theta^a, \label{wa02} \\
{\varpi^0}_0 & := s^* {\omega^0}_0 = - \we_a \theta^a, \label{w002} \\
{\varpi^a}_b & := s^* {\omega^a}_b = {\theta^a}_\mu d{e^\mu}_b - \eta^{ac}\eta_{bd} \we_c \theta^d + \we_b \theta^a + {\theta^a}_\mu \Gamma^\mu_{\nu\lambda} {e^\nu}_b dx^\lambda, \label{wab2} \\
{\varpi^0}_b & := s^* {\omega^0}_b = d\we_b - \we_c {\theta^c}_\mu d{e^\mu}_b - \we_c\theta^c \we_b - \we_c {\theta^c}_\mu \Gamma^\mu_{\nu\lambda} {e^\nu}_b dx^\lambda \label{w0b2} \\
& \qquad\qquad\qquad  + \half \we_c \eta^{cd}\we_d \eta_{ba} \theta^a + {e^\mu}_b P_{\mu\lambda} dx^\lambda \nonumber,
\end{align}
\end{subequations}
with $\Upsilon = \Upsilon_\mu dx^\mu$, $\Upsilon_b = \Upsilon_\mu {e^\mu}_b$, and $\we_a = e_a+\Upsilon_a$, where we have used that $d \Upsilon = 0$.

\subsubsection{First dressing}

The goal of the dressing field method is to erase the gauge symmetries present in the (pull-back of the) connection as given in \eqref{local_conf_connection}. Typically, the order one erases these degrees of freedom follows the decomposition \eqref{decomp_H} as (semi direct) products of the group itself, starting with the outer component. For this first dressing, we will hence consider removing the gauge freedom associated to the special conformal transformations, belonging to the subgroup $K$.

Even though the dressing method is not a gauge transformation, it is useful to look at first these transformations to get a hint of how to define them. Let us thus take a gauge element $\gamma_1 \in \mathcal{K}$, 
\begin{equation}
\label{expr_gamma1}
\gamma_1 = \left(\begin{array}{ccc}
1 & r_a & r_a \eta^{ab} r_b \\
& 1 & \eta^{ac} r_c \\
& & 1
\end{array}\right),
\end{equation}
and consider its action on the local connection, $\varpi^{\gamma_1} := \gamma_1^{-1} \varpi \gamma_1 + \gamma_1^{-1} d\gamma_1$. Using the matrix representation~\eqref{matrix_rep}, we have the transformations for the components of the connection,
\begin{subequations}
\label{trsf_conf_conn}
\begin{align}
\left({\varpi^a}_0\right)^{\gamma_1} & = {\varpi^a}_0 \label{trsf_wa0}\\
\left({\varpi^0}_0\right)^{\gamma_1} & = {\varpi^0}_0 - r_c {\varpi^c}_0 \label{trsf_w00} \\
\left({\varpi^a}_b\right)^{\gamma_1} & = {\varpi^a}_b + {\varpi^a}_0 r_b - r^a \eta_{bc} {\varpi^c}_0 \label{trsf_wab} \\
\left({\varpi^0}_b\right)^{\gamma_1} & = {\varpi^0}_b - r_c {\varpi^c}_b + \half r^2 \eta_{bc} {\varpi^c}_0 + {\varpi^0}_0 r_b - r_c{\varpi^c}_0 r_b + d r_b \label{trsf_w0b}
\end{align}
\end{subequations}

From \eqref{trsf_wa0} and \eqref{wa02}, one finds the invariance, 
\begin{equation}
\label{trsf_theta}
\left(\theta^a\right)^{\gamma_1} = \theta^a,
\end{equation}
and from \eqref{trsf_w00}, \eqref{w002}, and \eqref{trsf_theta}, we have,
\begin{equation}
\label{trsf_w00_2}
- \left({\varpi^0}_0\right)^{\gamma_1} = \left(e_a\right)^{\gamma_1} \theta^a + \left(\Upsilon_a\right)^{\gamma_1} \theta^a = e_a \theta^a + \Upsilon_a \theta^a + r_a \theta^a,
\end{equation}
meaning that $(e_a + \Upsilon_a)^{\gamma_1} = e_a + \Upsilon_a + r_a$.

Now, some authors already considered gauge fixing in the conformal connection, see \eg \cite{Sharpe96}, by putting the component ${\varpi^0}_0$ of the local connection to zero. Let us see how it is implemented in this framework. The equation $\left({\varpi^0}_0\right)^{\gamma_1} = 0$ will give us a hint on how to write the upcoming dressing field. Solving this equation, we find $\gamma_1$, as defined in \eqref{expr_gamma1}, to be parametrized by,
\begin{equation}
r_a = -e_a - \Upsilon_a.
\end{equation}

Hence, we define a dressing field $u_1$, which is a map $u_1 : U \subset M \rightarrow K$, with the defining equivariance $u_1^{\gamma_1} = \gamma_1^{-1} u_1$, for $\gamma_1$ an arbitrary element in the gauge group $\mathcal{K}_1$. We write it, with $\we_a = e_a+\Upsilon_a$, as,
\begin{equation}
u_1 = \left(\begin{array}{ccc}
1 & -\we_a & \we_a \eta^{ab} \we_b \\
& 1 & -\eta^{ac} \we_c \\
& & 1
\end{array}\right),
\end{equation}

One can then defines the dressed connection $\varpi_1 := u_1^{-1} \varpi u_1 + u_1^{-1} du_1$, invariant under gauge transformations $\gamma_1 \in \mathcal{K}_1$. We find,
\begin{equation}
\varpi_1 = \left(\begin{array}{ccc}
0 & {{\varpi_1}^0}_b & 0 \\
{{\varpi_1}^a}_0 & {{\varpi_1}^a}_b & \eta^{ac} {{\varpi_1}^0}_c \\
0 & \eta_{cb} {{\varpi_1}^c}_0 & 0
\end{array}\right),
\end{equation}
with,
\begin{subequations}
\begin{align}
{{\varpi_1}^a}_0 & = \theta^a \\
{{\varpi_1}^a}_b & = {\theta^a}_\mu d{e^\mu}_b + {\theta^a}_\mu \Gamma^\mu_{\nu\lambda} {e^\nu}_b dx^\lambda \\
{{\varpi_1}^0}_b & = {e^\mu}_b \Gamma_{\mu\lambda} dx^\lambda
\end{align}
\end{subequations}

\subsubsection{Second dressing}

Now, as we have mentioned before, any element in $\CO(n, 2)$ can once again be decomposed as the product of two elements, one in the Weyl subgroup $W = \R\setminus\{0\}$ and one in the Lorentz subgroup $O(n-1, 1)$. Consider the action of a gauge element $\gamma_0 = \left(\begin{matrix}
1 & 0 & 0 \\
0 & {S^a}_b & 0 \\
0 & 0 & 1
\end{matrix}\right) \in \mathcal{O}(n-1, 1)$ on the original local connection $\varpi$. We have,
\begin{equation}
\varpi^{\gamma_0} = \left(\begin{matrix}
{\varpi^0}_0 & {\varpi^0}_c {S^c}_b & 0 \\
{{S^{-1}}^a}_c {\varpi^c}_0 & {{S^{-1}}^a}_c {\varpi^c}_d S^d_b + {{S^{-1}}^a}_c d{S^c}_b & {{S^{-1}}^a}_c \eta^{cd} {\varpi^0}_d \\
0 & \eta_{cd} {\varpi^c}_0 {S^d}_b & - {\varpi^0}_0
\end{matrix}\right).
\end{equation}

We find, notably, $\left(\theta^a\right)^S = {{(S^{-1})}^a}_c \theta^c$. Since ${{\varpi^0}_0}^{\gamma_0} = {\varpi^0}_0 = - \we_a \theta^a$, we find also ${\we_a}^{\gamma_0} = \we_b {S^b}_a$, which means that we have the compatibility condition between consecutive dressing fields: ${u_1}^S = S^{-1} u_1 S$, as mentioned in the section we reviewed the dressing field method, \ref{ss:review_dressing}. Let us now define the dressing field $u_0$,
\begin{equation}
u_0 = \left(\begin{matrix}
1 & 0 & 0 \\
0 & {\theta^a}_\mu & 0 \\
0 & 0 & 1
\end{matrix}\right),
\end{equation}
which is such that ${u_0}^S = S^{-1} u_0$, and, from \eqref{trsf_wa0}, ${u_0}^{\gamma_1} = u_0$. The dressed connection $\varpi_0 := {u_0}^{-1} \varpi_1 u_0 + {u_0}^{-1} du_0$ is then, with ${u_0}^{-1} = \left(\begin{matrix}
1 & 0 & 0 \\
0 & {e^\mu}_a & 0 \\
0 & 0 & 1
\end{matrix}\right)$,
\begin{equation}
\varpi_0 = \left(\begin{matrix}
0 & P_{\nu\lambda} & 0 \\
\delta^\mu_\lambda & \Gamma^\mu_{\nu\lambda} & P^\mu_\lambda \\
0 & g_{\nu\lambda} & 0
\end{matrix}\right) dx^\lambda ,
\end{equation}
with $\Gamma$ and $P$ respectively the Christoffel symbols and the Schouten tensor associated to the metric $g$. This form of the conformal connection has already been obtained by dressing in \cite{FrancoisLM15,TheseJordan}, although not in the 2-frame parametrization. This means that they identified the remaining objects, \ie the Christoffel symbols and the Schouten tensor \textit{a posteriori}, while they are present from the beginning in this framework, and the dressing field method merely uncovers them.

\subsubsection{Conformal tractors}

Let us quickly recall the results of \cite{AttardF18} in which the (conformal) tractor bundle~\cite{BaileyEG94} is recovered through the dressing field method as a top-down construction. We will later use this same method to recover the projective tractor bundle. 

\smallskip
The representation space for $G_\cC$ and $H_\cC$ being $\R^{n+2}$, take sections $\varphi : U \subset M \rightarrow \R^{n+2}$ of the associated bundle $E_\cC := P_\cC \times_{H_\cC} \R^{n+2}$, parametrized as,
\begin{equation}
\varphi = \left(\begin{matrix}
\rho \\
l^\mu \\
\sigma
\end{matrix}\right).
\end{equation}

The authors of \cite{AttardF18} show that after reducing this associated bundle to ${E_\cC}_0$ by dressing these sections, \ie by defining $\varphi_1 := u_1^{-1} \varphi$ and $\varphi_0 := u_0^{-1} \varphi_1$, as well as defining the covariant derivative as $D_0 \varphi_0 := d \varphi_0 + \varpi_0 \varphi_0$ on ${E_\cC}_0$, one finds that the \emph{residual} transformations, which are the left over Weyl dilations that we could not reduce, imply a transformation law of the sections,
\begin{equation}
\varphi_0^Z  = \left(\begin{matrix}
z^-1 \left( \rho_0 - \gamma_\mu l_0^\mu + \frac{\sigma}{2} \gamma^2\right) \\[2mm]
z^{-1} \left(l_0^\mu - g^{\mu\nu} \gamma_\nu \sigma_0\right) \\[2mm]
z \sigma_0
\end{matrix}\right),
\end{equation}
with $\gamma_\mu = \partial_\mu \ln z$. These sections are then tractors and ${E_\cC}_0$ is the tractor bundle.

\subsection{The dressing field method applied to projective geometry}
\label{ss:proj}

The process is very similar to the case of conformal geometry treated in Section~\ref{ss:conf}, and we will follow hereafter the same plan. 

\subsubsection{Coefficients and first dressing}

First, we need to rewrite the coefficients of the Cartan connection in the matrix representation \eqref{proj_connection} by taking into account that in these expressions the coefficients $\Pi^\mu_{\nu\lambda}$ and $\Pi_{\mu\lambda}$ are related to, respectively, the Christoffel symbols and the Schouten tensor by a projective transformation as to render the first set of coefficients traceless \cite{KobayashiN64,Cartan37}. Hence, we use the well-known projective relations $\Pi^\mu_{\nu\lambda} = \Gamma^\mu_{\nu\lambda} + \delta^\mu_\nu \Upsilon_\lambda + \delta^\mu_\lambda \Upsilon_\nu$ with $\Upsilon_\mu = - \frac{1}{n+1} \Gamma^\rho_{\rho\mu}$ and $\Pi_{\mu\nu} = P_{\mu\nu} + \nabla_\mu \Upsilon_\nu - \Upsilon_\mu \Upsilon_\nu$, to define the Christoffel symbols $\Gamma$ of the projective structure and the projective Schouten tensor $P$ introduced in~\cite{GoverM17}. 
As in the conformal case, $\Pi^\mu_{\nu\lambda}$ measures the discrepancy between an arbitrary $2$-frame to be a projective frame.
It is straightforward to check that we have indeed $\Pi^\mu_{\mu\nu} = 0$ with these transformations. Then, we write the connection locally through a section $s$, where we shall write $\varpi := s^* \omega$ from now on, and we have,
\begin{equation}
\varpi := s^* \omega = \left(\begin{matrix}
\varpi^a_b & \varpi^a_0 \\
\varpi^0_b & \varpi^0_0
\end{matrix}\right),
\end{equation}
with,
\begin{subequations}
\label{local_proj_conn}
\begin{align}
{\varpi^a}_0 & = \theta^a, \label{proj_wa02} \\
{\varpi^0}_0 & = - \left(e_a + \Upsilon_a\right) \theta^a \label{proj_w002} \\
{\varpi^a}_b & = {\theta^a}_\mu d{e^\mu}_b + \left(e_b + \Upsilon_b\right) \theta^a + {\theta^a}_\mu \Gamma^\mu_{\nu\lambda} {e^\nu}_b dx^\lambda, \label{proj_wab2} \\
{\varpi^0}_b & = d\left(e_b+\Upsilon_b\right) - \left(e_c+\Upsilon_c\right) {\theta^c}_\mu d{e^\mu}_b - \left(e_c+\Upsilon_c\right)\theta^c \left(e_b+\Upsilon_b\right) + {e^\mu}_b P_{\mu\lambda} dx^\lambda \label{proj_w0b2} \\
& \quad - \left(e_c+\Upsilon_c\right) {\theta^c}_\mu \Gamma^\mu_{\nu\lambda} {e^\nu}_b dx^\lambda \nonumber,
\end{align}
\end{subequations}
where $d$ is now the differential of $M$.

\smallskip
Similarly to the conformal case, we start by reducing the gauge symmetries associated to the outermost subgroup in the semi direct product decomposition \eqref{proj_decomp_h}, \ie ${\R^n}^*$. Let us first consider the action of a gauge transformation $\gamma_1 = \left(\begin{matrix}
\delta^a_b & 0 \\
r_b & 1
\end{matrix}\right)$. We have, for $\varpi^{\gamma_1} = \gamma_1^{-1} \varpi \gamma_1 + \gamma_1^{-1} d\gamma_1$,
\begin{equation}
\varpi^{\gamma_1} = \left(\begin{matrix}
{\varpi^a}_b + r_a {\varpi^b}_0 & {\varpi^a}_0 \\
{\varpi^0}_b - {\varpi^c}_b r_c + {\varpi^0}_0 r_b - r_c {\varpi^c}_0 r_b - dr_b & {\varpi^0}_0 - r_c {\varpi^c}_0
\end{matrix}\right).
\end{equation}

Since the computations are very similar to the conformal case, especially since that ${\theta^a}^{\gamma_1} = \theta^a$ and $\we_a^{\gamma_1} := (e_a + \Upsilon_a)^{\gamma_1} = \we_a + r_a$, the dressing field $u_1 : U \subset M \rightarrow \R^n$ is naturally,
\begin{equation}
\label{proj_u1}
u_1 = \left(\begin{matrix}
1& 0 \\
- \we_a & 1
\end{matrix}\right),
\end{equation}
so that $u_1^{\gamma_1} = \gamma_1^{-1} u_1$. One then defines the dressed connection $\varpi_1 := u_1^{-1} \varpi u_1 + u_1^{-1} du_1$ to find,
\begin{equation}
\varpi_1 = \left(\begin{matrix}
{\theta^a}_\mu {de^\mu}_b + {\theta^a}_\mu \Gamma^\mu_{\nu\lambda} {e^\nu}_b dx^\lambda & \theta^a \\
{e^\mu}_b P_{\mu\nu} dx^\nu & 0
\end{matrix}\right).
\end{equation}

The (local) connection $\varpi_1$ is invariant under the gauge transformations $\gamma_1$ spawned by ${\R^n}^*$.

\subsubsection{Second dressing}

The aim of the second dressing is to render the connection invariant under the gauge transformations associated to the remaining symmetries, other than rescaling, which corresponds here to those spawned by $\GL(n, \R)$, \ie of the form $\gamma_0 = \left(\begin{matrix}
{S^a}_b & 0 \\
0 & 1
\end{matrix}\right)$. The dressing field $u_0$ has to be such that $u_0^{\gamma_0} = \gamma_0^{-1} u_0$ and $u_0^{\gamma_1} = u_0$, which are the compatibility conditions for consecutive dressing fields, together with ${u_1}^{\gamma_0} = \gamma_0^{-1} u_1 \gamma_0$, which is verified. The dressing field $u_0$ is given by,
\begin{equation}
\label{proj_u0}
u_0 = \left(\begin{matrix}
{\theta^a}_\mu & 0 \\
0 & 1
\end{matrix}\right).
\end{equation}

This then leads to the following dressed connection, defined as $\varpi_0 := u_0^{-1} \varpi_1 u_0 + u_0^{-1} du_0$,
\begin{equation}
\label{proj_omega0}
\varpi_0 = \left(\begin{matrix}
\Gamma^\mu_{\nu\lambda} & \delta^\mu_\lambda \\
P_{\nu\lambda}& 0
\end{matrix}\right) dx^\lambda,
\end{equation}
with, respectively, $\Gamma$ and $P$ the projective Christoffel symbols and the projective Schouten tensor \cite{GoverM17}. The only gauge transformations left acting on the connection $\varpi_0$ are the ones associated to the dilations, Weyl rescalings. We will compute the action of such rescalings shortly.

\subsubsection{Projective tractors}
Tractors exist not only for conformal calculus, but also for projective calculus \cite{BaileyEG94,GoverM17}. We will see now that, just like for conformal tractors (see \cite{AttardF18} or section \ref{ss:conf}), the dressing field method is a way to construct the projective tractor bundle.

Here, the defining representation space for the groups $G_\cP$ and $H_\cP$ is $\R^{n+1}$. Hence, we take sections $\varphi : U \subset M \rightarrow \R^{n+1}$ of the associated vector bundle $E_\cP = P_\cP \times_{H_\cP} \R^{n+1}$, parametrized as,
\begin{equation}
\varphi = \left(\begin{matrix}
l^a \\
\sigma
\end{matrix}\right),
\end{equation}
and define the covariant derivative so that $D \varphi := d\varphi + \varpi \varphi$, with $\omega$ the original projective Cartan connection computed in section \ref{s:proj_connection}.

The dilation symmetry $Z \in \mathcal{W} = \R \setminus \{0\}$ leads to the transformation law,
\begin{equation}
\varphi^Z = Z^{-1} \varphi = \left(\begin{matrix}
l^a \\
z^{-1} \sigma
\end{matrix}\right).
\end{equation}

However, just like in the conformal case, $E_\cP$ is not the true tractor bundle. One should reduce the bundle $E_\cP$ by dressing both the projective tractor and the connection, with the dressing fields $u_1$ \eqref{proj_u1} and $u_0$ \eqref{proj_u0}, to obtain $\varphi_1 \in {E_\cP}_1 := E_\cP^{u_1}$, then $\varphi_0 \in {E_\cP}_0 := {E_\cP}_1^{u_0}$, and the dressed connection $\varpi_0$ and then look at the transformation laws of these two objets on ${E_\cP}_0$. We have, by definition, $\varphi_1 := u_1^{-1} \varphi = \left(\begin{matrix}
l^a \\
\sigma + \we_a l^a
\end{matrix}\right)$ and $\varphi_0 := u_0^{-1} \varphi_1 = \left(\begin{matrix}
{e^\mu}_a l^a \\
\sigma + \we_a l^a
\end{matrix}\right) =: \left(\begin{matrix}
l^\mu \\
\sigma_0
\end{matrix}\right)$.

The covariant derivative on ${E_\cP}_0$ will be $D_0 \varphi_0 := d\varphi_0 + \varpi_0 \varphi_0$, or in components,
\begin{equation}
{D_0}_\nu \varphi_0 = \left(\begin{matrix}
\nabla_\nu l^\mu + \sigma \delta^\mu_\nu \\[2mm]
\nabla_\nu \sigma + P_{\nu\lambda} l^\lambda
\end{matrix}\right),
\end{equation}
which is the same expression that was defined in \cite{BaileyEG94,GoverM17}. As we will see next, ${E_\cP}_0$ is the projective tractor bundle as defined in \cite{BaileyEG94,GoverM17}.

\subsubsection{Projective transformations of the connection and tractors}

In the conformal case, the tractor connection and the tractors themselves have specific transformation laws. These transformations can be recovered through the dressing field method, where they are spawned by the remaining gauge symmetry that could not be reduced, dilations. Here, we will apply the same method in the case of projective geometry, to see what transformation arise from the remaining gauge symmetry.

The structure group of the original principal bundle was $H_\cP = \left(\GL(n, \R) \times \R \setminus \{0\}\right) \ltimes {\R^n}^*$. We have first reduced the gauge transformations due to ${\R^n}^*$ and then those due to $\GL(n, \R)$. We are then left with $\R \setminus \{0\}$-gauge transformations, which correspond to dilations. In the matrix decomposition of the structure group $H_\cP$, this corresponds to elements of the form,
\begin{equation}
Z = \left(\begin{matrix}
\delta^a_b & 0 \\
0 & z
\end{matrix}\right).
\end{equation}

Let us first compute the gauge transformation, associated to dilations, of the original normal Cartan connection. We have $\varpi^Z := Z^{-1} \varpi Z + Z^{-1} dZ = \left(\begin{matrix}
{\varpi^a}_b & z {\varpi^a}_0 \\
z^{-1} {\varpi^0}_b & {\varpi^0}_0 + z^{-1} dz
\end{matrix}\right)$. Using the definitions of the coefficients of the connection \eqref{proj_wa02} and \eqref{proj_w002}, we immediately find the transformation ${\theta^a}^Z = z \theta^a$, which leads to $\we_a^Z = z^{-1} \left(\we_a - \zeta_\mu {e^\mu}_a\right)$, for $\zeta_\mu = z^{-1} \partial_\mu z$. Then, solving $({\varpi^a}_b)^Z = {\varpi^a}_b$ and $({\varpi^0}_b)^Z = z^{-1} {\varpi^0}_b$ for the expressions \eqref{proj_wab2} and \eqref{proj_w0b2} we find the following transformations,
\begin{subequations}
\begin{align}
\left(\Gamma^\mu_{\nu\lambda}\right)^Z & = \Gamma^\mu_{\nu\lambda} + \delta^\mu_\lambda \zeta_\nu + \delta^\mu_\nu \zeta_\lambda, \label{proj_trsf_christo} \\
\left(P_{\mu\lambda}\right)^Z &  = P_{\mu\lambda} + \nabla_\lambda \zeta_\mu - \zeta_\lambda \zeta_\mu, \label{proj_trsf_schouten}
\end{align}
\end{subequations}
which are, by definition, the relations describing a projective transformation acting on the Christoffel symbols and on the projective Schouten tensor, respectively.

\smallskip
Instead of building the dressing in two steps, one could have performed it in one step. Take the two dressing fields $u_1$ \eqref{proj_u1} and $u_0$ \eqref{proj_u0} and define $u = u_1 u_0 = \left(\begin{matrix}
{\theta^a}_\mu & 0 \\
- \we_a {\theta^a}_\mu & 1
\end{matrix}\right)$ the dressing field transforming directly the undressed connection $\omega$ into $\varpi_0 := u^{-1} \varpi u + u^{-1} du$ \eqref{proj_omega0}. According to the transformations under the gauge symmetry $Z$ we found in the previous paragraph, we have
\begin{equation}
\label{proj_trsf_u}
u^Z = \left(\begin{matrix}
z {\theta^a}_\mu & 0 \\
- \left(\we_a - \zeta_\rho {e^\rho}_a\right) {\theta^a}_\mu & 1
\end{matrix}\right).
\end{equation}

Now, we need this transformation law of $u$ \eqref{proj_trsf_u} to be of the form $u^Z = Z^{-1} u C(z)$ so that the transformation of the tractors $\varphi_0 := u^{-1} \varphi \in {E_\cP}_0$ is given by $\varphi_0^Z = u^Z \varphi^Z = \left(Z^{-1} u C(z)\right)^{-1} \left(Z^{-1} \varphi\right) = C(z)^{-1} \varphi_0$ and similarly for the connection $\varpi_0^Z = C(z)^{-1} \varpi_0 C(z) + C(z)^{-1} d C(z)$ \cite{AttardF18}. Since $Z^{-1}u = \left(\begin{matrix}
{\theta^a}_\mu & 0 \\
- z^{-1} \we_a {\theta^a}_\mu & z^{-1}
\end{matrix}\right)$, we readily find that $C(z) = z \left(\begin{matrix}
 \delta^\mu_\nu & 0 \\
\zeta_\nu & 1
\end{matrix}\right)$. Hence, we have the residual transformations laws,
\begin{equation}
\varphi_0^Z = \left(\begin{matrix}
z^{-1} \delta^\mu_\nu & 0 \\
-z^{-1} \zeta_\nu & z^{-1}
\end{matrix}\right)
\left(\begin{matrix}
l^\nu \\
\sigma_0
\end{matrix}\right)
=
\left(\begin{matrix}
z^{-1} l^\mu \\
z^{-1} \left(\sigma_0 - l^\nu \zeta_\nu\right)
\end{matrix}\right),
\end{equation}
and,
\begin{equation}
\varpi_0^Z = \left(\begin{matrix}
\Gamma^\mu_{\nu\lambda} + \delta^\mu_\nu \zeta_\lambda + \delta^\mu_\lambda \zeta_\nu & \delta^\mu_\lambda \\[1mm]
P_{\nu\lambda} + \nabla_\mu \zeta_\nu - \zeta_\mu \zeta_\nu & 0
\end{matrix}\right) dx^\lambda\,.
\end{equation}

We see that we recover the projective transformations for the Christoffel symbols and the projective Schouten tensor in the connection, just like in \eqref{proj_trsf_christo} and \eqref{proj_trsf_schouten}, and we obtain as a bonus of this top-down construction the transformation of the projective tractors.

\section{Conclusion}

We have seen how the \emph{dressing field method} is used to reduce the gauge symmetries of conformal and projective connections, using the framework of 2-frame bundles, effectively going from normal Cartan connections to mostly gauge invariant local connections. While the pedagogical case of Lorentz geometry does not feature any remaining gauge symmetry, both the conformal and the projective geometries feature a remaining 1-dimensional symmetry, associated to dilations. Though the example of application of the dressing field method to conformal structures was already known, albeit not in the language of 2-frame bundles, we have seen how to apply it to projective structures. 

While the method certainly looks like a \emph{dressing} in the matrix representation \cite{TheseJordan}, where components of the connections are only identified \textit{a posteriori}, here in the context of 2-frame bundles, where geometrical objects are already present at the level of Cartan connections, the method looks more like an \emph{undressing} upon using frames pertaining to $G$-structures. Indeed, objects of interest are identified \textit{a priori} and are uncovered by the process. For example for conformal and projective connections, we get rid of every terms depending on frame coordinates in the connection, but the Christoffel symbols and the Schouten tensor for the metric as introduced in~\eqref{realisation_metric}. 

Lastly, it is known that one can obtain the conformal tractor bundle from the dressing field method on a conformal structure \cite{AttardF18}. We have shown here how to construct the projective tractor bundle, meeting its definition in \cite{BaileyEG94}, with the help of that method on projective structures considered as a $G$-structure. 

Hence, the dressing field method is worth studying, as it provides a constructive way to reduce Cartan connections on any kind of fiber bundle, and to define associated tractors. More examples may follow, with unusual geometry, where defining tractors the historical way is not straightforward, but should be easier with the dressing field method. For instance, tractors for Newton-Cartan and Bargmann geometries will be treated along this line in a forthcoming paper.

\subsection*{Acknowledgements}

The project leading to this publication has received funding from the Excellence Initiative of Aix-Marseille University - A*Midex, a French ``Investissements d’Avenir programme'' AMX-19-IET-008 and AMX-19-IET-009.

\appendix

\section{Some explicit calculation \label{sect:appendix}}
\setcounter{equation}{0}
\renewcommand\theequation{\Alph{section}.\arabic{equation}}

In this appendix, one performs some computational steps along the line proposed in the seminal papers \cite{KobayashiN64,Ogiue67} in order to compute explicitly the local expressions of the Cartan connection $\womega$ over the reduced $2$-frame bundle. For the sake of completeness, this will be done for the conformal case in order to recover expressions given in~\cite[see §7]{Ogiue67} and recast in \eqref{eq:normal-Cartan} in frame coordinates.
One must compute the standard relation at any point $p\in P$ and trivialized by $p=\sigma(x)\cdot h$, where $x=\pi(p)$,
\begin{equation}
\label{eq:Cartan-gauge}
\womega_{|\sigma\cdot h} = \text{Ad}(h^{-1}) \womega_{|\sigma} + h^{-1} dh
\end{equation}
given a trivializing section $\sigma$ of the 2-frame bundle and $h\in H_{\cal C}$.  The local expression of the Cartan form is thus given by two pieces, one given by the adjoint representation and the other one coming from the Maurer-Cartan form. Let us elaborate on these two.

\subsection{The adjoint representation}

Here, the adjoint representation of $H_{\cal C}$  as a subgroup of $G_{\cal C}$ (cf \eqref{def_G_matrix}) acts on the graded Lie algebra $\mathfrak{g}_{\cal C} = \mathfrak{g}_{-1} \oplus \mathfrak{g}_0 \oplus \mathfrak{g}_1$ must be computed through the jet composition law. This requires the use of jets at $0$, $h_3 = j_3(0)$, up to order $3$ of the transformations given in \eqref{action_mobius_rn}; namely, beside the 1st and 2nd orders respectively given in \eqref{expr_1st_order} and \eqref{rel_2nd_order}, one must add the defining condition for the 3rd order jets
\begin{equation}
\label{rel_3rd_order}
h^k{}_{ijm} := \oint_{ijm} \big(
2 {h^k}_i h_j h_m - {h^k}_r \eta^{rs} h_s \eta_{ij} h_m - \frac{1}{2}\, \eta^{rs}h_r h_s \eta_{ij} {h^k}_m \big)
\end{equation}
where $\oint_{ijm}$ means summation over the cyclic permutation of the three indices $i,j,m$.

According to \cite{Kobayashi61} (see also \cite{TheseGrasseau}), one can write
\[
\Ad(h_3) A = \left.\dfrac{d}{dt} j_3(h\circ f_t \circ h^{-1}) \right|_{t=0}, \qquad A =  \left.\dfrac{d}{dt} j_3(f_t)(0) \right|_{t=0}, \qquad f_{t=0} = \text{Id}_{\R^n},
\]
where $f_t$ is a flow generated by $A=(A^k,A^k{}_\ell,A_\ell)\in \mathfrak{g}_{\cal C} = \mathfrak{g}_{-1} \oplus \mathfrak{g}_0 \oplus \mathfrak{g}_1$ where the last components are given by linearizing \eqref{rel_2nd_order} 
\begin{align}
\label{eq:g1}
A^k{}_{\ell m} = \eta^{kr}A_r\eta_{\ell m} - \delta^k_\ell A_m - \delta^k_m A_\ell \quad \Rightarrow A_m = -\frac{1}{n} A^r{}_{rm}.
\end{align}
Explicitly, one gets for the components of $B = \Ad(h_3) A \in \mathfrak{g}_{\cal C}$ according to the graduation, and setting $\bar{h} = h^{-1}$,
\begin{align}
\label{eq:adj}
B^k &= {h^k}_\ell A^\ell \notag\\
{B^k}_\ell &= {h^k}_r {A^r}_s \bar{h}^s{}_\ell + {h^k}_{rs}A^r\bar{h}^s{}_\ell \\
{B^k}_{\ell m} &=  h^k{}_{rst} A^t \bar{h}^s{}_\ell \bar{h}^r{}_m +
h^k{}_{rs} {A^s}_t \bar{h}^t{}_\ell \bar{h}^r{}_m +
h^k{}_{rs} {A^r}_t \bar{h}^s{}_\ell \bar{h}^t{}_m +
h^k{}_r {A^r}_s \bar{h}^s{}_{\ell m} \notag\\
& \quad +
h^k{}_{rs} A^r \bar{h}^s{}_{\ell m} +
h^k{}_r {A^r}_{st} \bar{h}^s{}_\ell \bar{h}^t{}_m \notag
\end{align}
in which expressions \eqref{rel_2nd_order} and \eqref{rel_3rd_order} must be substituted at the end. One sees the occurrence of the 3rd order jet of $h$. 
In addition, the inverse jet of $h_3$ is given by
\begin{align}
\bar{h}^k{}_r h^r{}_\ell &= \delta^k_\ell, \qquad \text{with}\quad \eta_{k\ell} {h^k}_i {h^\ell}_j = z^{-2} \eta_{ij} \notag\\[-4mm]
& \label{eq:inverse-jet}\\
\bar{h}^k{}_{\ell m} &= - \bar{h}^k{}_r h^r{}_{st} \bar{h}^s{}_\ell \bar{h}^t{}_m \notag 
\end{align}
where of course, according to the jet group law for $H^3_{\cal C}(n)$, $\bar{h}_3$ fulfills both the prolongation \eqref{rel_2nd_order} and \eqref{rel_3rd_order}; it is worthwhile to give the relation 
\begin{align}
\label{eq:G1-inverse}
\bar{h}_j = - \frac{1}{n} h^r{}_s \bar{h}^s{}_{rj} = - h_r \bar{h}^r_j
\end{align}
where equations \eqref{eq:inverse-jet} and \eqref{eq:paramG1} have been used.

Owing to \eqref{eq:g1}, let us now compute the free parameters in $\mathfrak{g}_1 \simeq (\R^n)^*$ parametrizing the 2nd order jets given in~\eqref{eq:adj}
\begin{align*}
-n B_m & = B^\ell{}_{\ell m} \\
& = \big( \bar{h}^s{}_\ell h^\ell{}_{rst} A^r - n h_r A^r{}_t + \bar{h}^s{}_\ell h^\ell{}_{rt} A^r{}_s -n A_t \big) \bar{h}^t{}_m + \bar{h}^s{}_{\ell m} \left( h^\ell{}_{rs} A^r + h^\ell{}_r A^r{}_s\right)
\end{align*}
where \eqref{eq:paramG1} and \eqref{eq:g1} have been already used. Next, by \eqref{rel_3rd_order} and \eqref{eq:inverse-jet}, and after a straightforward but lengthy computation one ends with
\begin{align}
\label{eq:adj-g1}
B_m = \big( A_t + h_r A^r{}_t - h_r A^r h_t + \frac{1}{2} \eta^{ij} h_i h_j \eta_{rt} A^r \big) \bar{h}^t{}_m \ .
\end{align}
This will avoid us to compute the 3rd order jet of $h^{-1}$.

\subsection{The Maurer-Cartan form on \texorpdfstring{$H^3(n)$}{H3(n)}}

Usually, the Maurer-Cartan form is given by $\Theta_{\text{MC}} (X) (g) = T_g L_{g^{-1}} \cdot X_{|g}$, for any $X\in \text{Vect}(G)$.
One must compute in the sense of jet composition for $H_{\cal C}$ the differential map
\[
\left. \dfrac{d}{dt} j_2 \big( h^{-1}\circ\gamma(t)\big) \right|_{t=0} = T_h L_{h^{-1}} \cdot X_{|h}  \in T_e H_{\cal C} = \mathfrak{h}_{\cal C}
\]
for any curve $\gamma(t)$ in $H^2_{\cal C}(n)$ such that $\gamma(0) = h$ and $\dot{\gamma}(0) =X_{|h}$.
Thus, computing the jet at $0$ \[
\left.\dfrac{d}{dt} j_2 \big(h^{-1} \circ \gamma(t)\big)(0)\right|_{t=0}
\]
one gets for $h^{-1} d h = \bar{h} \circ dh$ according to the graduation $\mathfrak{h}_{\cal C} = \mathfrak{g}_0 \oplus \mathfrak{g}_1$
\begin{align}
\bar{h}^k{}_r X^r_\ell &\to \bar{h}^k{}_r dh^r{}_\ell \notag\\[-4mm] 
& \label{eq:MC-0}\\
\bar{h}^k_r X^r_{\ell m} + \bar{h}^k_{rs} (h^r_\ell X^s_m + h^r_m X^s_\ell) &\to \bar{h}^k{}_r dh^r_{\ell m} + \bar{h}^k_{rs} h^r{}_\ell dh^s{}_m + \bar{h}^k_{rs} h^r{}_m dh^s{}_\ell \ . \notag
\end{align}
Recalling that the 2nd order jet are defined by \eqref{rel_2nd_order}, the 2nd order jet of the inverse in the last equation must be replaced according to \eqref{eq:inverse-jet}. One gets after some algebra
\begin{align}
\label{eq:MC-1}
-\frac{1}{n} \big( \bar{h}^\ell{}_r dh^r_{\ell m} + \bar{h}^\ell_{rs} h^r{}_\ell dh^s{}_m + \bar{h}^\ell_{rs} h^r{}_m dh^s{}_\ell \big) =  dh_m - h_s \bar{h}^s{}_r d h^r{}_m
\end{align}
which gives the $\mathfrak{g}_1$-component of the Maurer-Cartan form on $H^2_{\cal C}(n)$, see \eqref{eq:g1}.

\subsection{Computation of the local expression for the Cartan connection}

Exchanging $h\leftrightarrow \bar{h}$ and replacing $A$ by $\womega_{|\sigma(x)} = (\sigma^*\womega)(x)$ in the equations \eqref{eq:adj} and \eqref{eq:adj-g1} and then collecting terms according to the graduation $\mathfrak{g}_{-1} \oplus \mathfrak{g}_0 \oplus \mathfrak{g}_1$, one can compute \eqref{eq:Cartan-gauge}. For the components in $\mathfrak{g}_{-1} \oplus \mathfrak{g}_0$ one gets the local expressions:
\begin{align}
\womega^k &= \bar{h}^k{}_\ell\, \sigma^*\womega^\ell \notag\\[-4mm]
& \label{eq:-1et0} \\
\womega^k{}_\ell &= \bar{h}^k{}_r dh^r{}_\ell + \bar{h}^k{}_r\, \sigma^*\womega^r{}_s h^s{}_\ell + \bar{h}^k{}_{rs}\, \sigma^*\womega^r h^s{}_\ell  \notag\\
&= \bar{h}^k{}_r dh^r{}_\ell + \bar{h}^k{}_r\, \sigma^*\womega^r{}_s h^s{}_\ell - \eta^{kr} h_r  \eta_{\ell s} \womega^s + h_\ell \womega^k + \delta^k_\ell h_r \womega^r \notag
\end{align}
where, beside the Maurer-Cartan part, first formula in \eqref{eq:MC-0}, once more, the 2nd order jet for the inverse has been replaced according to \eqref{eq:inverse-jet} and the defining relation \eqref{rel_2nd_order} has been used. It remains to compute the component in $\mathfrak{g}_1$. 
This is achieved by combining both \eqref{eq:MC-1} and \eqref{eq:adj-g1}; one obtains
\[
\womega_m = dh_m - h_s \bar{h}^s{}_r d h^r{}_m + 
\big( \sigma^*\womega_t + \bar{h}_r \,\sigma^*\womega^r{}_t - \bar{h}_r \,\sigma^*\womega^r h_t + \frac{1}{2} \eta^{ij} \bar{h}_i \bar{h}_j \eta_{rt} \,\sigma^*\womega^r \big) h^t{}_m 
\] 
and then using \eqref{eq:G1-inverse} and $\eta_{k\ell} {h^k}_i {h^\ell}_j = z^{-2} \eta_{ij}$, one gets
\begin{align}
\womega_m &= dh_m - h_s \bar{h}^s{}_r d h^r{}_m + \sigma^*\womega_t h^t{}_m - h_\ell \bar{h}^\ell{}_r \,\sigma^*\womega^r{}_t h^t{}_m - h_\ell \bar{h}^\ell{}_r \,\sigma^*\womega^r h_m \notag\\
& \qquad+ \frac{1}{2} \eta^{ij}h_i h_j \eta_{mr} \bar{h}^r{}_s \,\sigma^*\womega^s \notag\\
&= dh_m - h_k \womega^k{}_m + h_k \womega^k h_m +  \sigma^*\womega_k h^k{}_m - \frac{1}{2} \eta^{ij} h_i h_j \eta_{mk} \womega^k
\label{eq:omega1}
\end{align}
if one wishes to use expressions previously obtained in \eqref{eq:-1et0}.

To sum up, the local expressions \eqref{eq:-1et0} and \eqref{eq:omega1} for the Cartan connection are those given in~\cite[see §7]{Ogiue67} and translated in frame coordinates \eqref{eq:normal-Cartan} with a specific choice of the trivializing section $\sigma: U \to P_{\cal C}, x \mapsto (x^\mu, \sigma^\mu{}_a, \sigma^\mu_{ab})$ and reconsidered as
\[
x \mapsto (x^\mu, \delta^\mu_\nu, - \Gamma^\mu_{\nu\rho}) , \qquad \text{with } \sigma^\mu{}_a \bar{\sigma}^a{}_\nu = \delta^\mu_\nu \text{ and } \Gamma^\mu_{\nu\rho} = \sigma^\mu_{ab} \bar{\sigma}^a{}_\nu \bar{\sigma}^b{}_\rho
\]
projecting onto the $1$-jet of the identity $(x^\mu, \delta^\mu_\nu)$.

The very recovering of expressions \eqref{eq:normal-Cartan} is made by inverting the trivialization $e_2 = \sigma(x) \cdot h$ for $x = \pi(e_2)$, namely,
\begin{align*}
\bar{h}^k{}_\ell &=  \theta^k{}_\mu \sigma^\mu{}_\ell \\
e_m &= h_m + \sigma_\ell h^\ell{}_m = h_m  + \sigma_\ell \bar{\sigma}^\ell{}_\rho e^\rho{}_m
\end{align*}
which leads to
\begin{align}
\womega^k &= \bar{h}^k{}_\ell\, \sigma^*\womega^\ell = \bar{h}^k{}_\ell \bar{\sigma}^\ell{}_\mu dx^\mu = \theta^k{}_\mu dx^\mu = \theta^k \notag\\
\womega^k{}_\ell &= \bar{h}^k{}_r dh^r{}_\ell + \bar{h}^k{}_r\, \sigma^*\womega^r{}_s h^s{}_\ell - \eta^{kr} h_r  \eta_{\ell s} \womega^s + h_\ell \womega^k + \delta^k_\ell h_r \womega^r \notag\\
&= {\theta^k}_\mu d{e^\mu}_\ell - \eta^{kr}\eta_{\ell s} e_r \theta^s + e_\ell \theta^k + \left(e_r \theta^r\right)\delta^k_\ell + {\theta^k}_\mu \left( \sigma^\mu{}_r \sigma^*\womega^r{}_s \bar{\sigma}^s{}_\nu + \sigma^\mu{}_r d \bar{\sigma}^s{}_\nu \right)\notag\\
&\qquad  + {\theta^k}_\mu ( g^{\mu\rho} \sigma_s \bar{\sigma}^s{}_\rho g_{\nu\lambda} - \delta^\mu_\lambda \sigma_s \bar{\sigma}^s{}_\nu - \delta^\mu_\nu \sigma_s \bar{\sigma}^s{}_\lambda ) dx^\lambda {e^\nu}_\ell\ . \notag
\end{align}
Then computing explicitly $\sigma^*\womega^r{}_s$ from the definitions
\eqref{def_womegaab} and \eqref{def_thetaab}, the last line simplifies and amounts to defining
$\Pi^\mu_{\nu\lambda} dx^\lambda = \sigma^\mu{}_r \, (\sigma^*\varphi^r{}_s) \bar{\sigma}^s{}_\nu$, a result which again shows that $\Pi^\mu_{\nu\lambda}$ depends on $x$ only. Hence, one recovers \eqref{def_wab}. 

Likewise, for the $\mathfrak{g}_1$-component and after replacing the $\mathfrak{g}_0$-component by its local coordinate expression as in \eqref{def_wab}, one gets back the local expression~\eqref{def_wb}
\begin{align*}
\womega_m & = de_m - e_r \womega^r_m + e_r \theta^r e_m + {e^\mu}_m \Pi_{\mu\nu} dx^\nu - \half \eta^{rs}\eta_{m\ell} e_r e_s \theta^\ell
\end{align*}
where after some cancellation $\Pi_{\mu\nu} dx^\nu = (\sigma^* \womega_k) \bar{\sigma}^k{}_\mu$
thus depending on~$x$ only.


\end{document}